\pdfoutput=1

\documentclass[11pt,a4paper]{article}

\usepackage{jcappub}
\usepackage[utf8]{inputenc}
\usepackage{amsmath}
\usepackage{amsthm}
\usepackage{amssymb}
\usepackage{enumitem}   
\usepackage[english]{babel}
\usepackage{url}
\usepackage{mathtools}
\usepackage{slashed}
\usepackage{multirow}
\usepackage{lipsum}
\usepackage{xcolor}
\usepackage{tcolorbox}
\usepackage{xfrac}

\begin{document}


\hfill MS-TP-26-21


\title{Signal-to-Noise Ratio Contours for LISA}

\author[a,b]{Kai~Schmitz}
\author[c]{and Joseph D.\ Romano}

\affiliation[a]{Institute for Theoretical Physics, University of M\"unster, 48149 M\"unster, Germany}
\affiliation[b]{Kavli IPMU (WPI), UTIAS, The University of Tokyo, Kashiwa, Chiba 277-8583, Japan}
\affiliation[c]{Department of Physics and Astronomy, UT Rio Grande Valley, Brownsville,TX 78520, USA}

\emailAdd{kai.schmitz@uni-muenster.de}
\emailAdd{joseph.romano@utrgv.edu}


\abstract{The Laser Interferometer Space Antenna (LISA) will search for a stochastic gravitational-wave (GW) background at millihertz frequencies, from both astrophysical and cosmological sources, and thereby open a new chapter in GW astronomy. In the literature, LISA's sensitivity to prospective GW background (GWB) signals is often quantified in terms of an expected signal-to-noise ratio (SNR) assuming perfect knowledge of the detector noise. The commonly employed expression for the SNR is, however, valid only in the limit of a weak GWB signal, which renders a large number of SNR values reported in the literature inaccurate. In this paper, we address this issue by deriving for the first time an expression for the expected optimal SNR of a LISA auto-correlation measurement that is valid at arbitrary signal strength. Based on our generalized expression, we conclude that LISA data worth an observing time of $T_{\rm obs}$ across the frequency band from $f_{\rm min}$ to $f_{\rm max}$ will never yield an SNR in excess of $\textrm{SNR}_{\rm max} = \sqrt{T_{\rm obs}(f_{\rm max}-f_{\rm min})}$, which evaluates to $\textrm{SNR}_{\rm max} \lesssim 10^4$ for typical mission parameters. We illustrate our findings in terms of generalized power-law-integrated (PLI) sensitivity curves at different SNR levels, i.e., LISA SNR contour lines in plots of the GW energy-density power spectrum. In contrast to earlier work on PLI sensitivity curves, we notably find that the LISA SNR contours are bounded from above, approximately by the LISA strain noise curve multiplied by a factor of Euler's number $e$. For GWB signals not much weaker than this range, the expected SNR for a LISA auto-correlation measurement needs to be evaluated based on our new expression. Our numerical results for the LISA SNR contours are available on Zenodo~\cite{zenodo}.}


\maketitle


\section{Introduction}
\label{sec:introduction}


The Laser Interferometer Space Antenna (LISA)~\cite{LISA:2017pwj,Baker:2019nia}, the first gravitational-wave (GW) laser interferometer experiment in space, is set to open a new window to the GW Universe. With an interferometer arm length of millions of kilometers, LISA will be sensitive to GWs in the millihertz frequency range, thus filling a crucial gap in the GW frequency spectrum between ground-based observations with LIGO, Virgo, and KAGRA at frequencies in the audio band and pulsar timing array (PTA) observations in the nanohertz band (see Ref.~\cite{Bailes:2021tot} for a review on GW astronomy in the 2020s and 2030s). The science program for LISA is exceptionally rich and covers a large range of topics in astrophysics~\cite{LISA:2022yao} and cosmology~\cite{LISACosmologyWorkingGroup:2022jok}. Next to a variety of astrophysical transients, LISA is also expected to be sensitive to GW background (GWB) signals, both from astrophysical~\cite{Chen:2018rzo,Babak:2023lro} and cosmological sources~\cite{Caprini:2024hue,Blanco-Pillado:2024aca,LISACosmologyWorkingGroup:2024hsc} (see Refs.~\cite{Regimbau:2011rp,Caprini:2018mtu} for reviews on these two types of sources, respectively).

LISA's sensitivity to GWB signals is often quantified in terms of an expected signal-to-noise ratio (SNR) for an appropriately constructed auto-correlation observable~\cite{Allen:1997ad,Maggiore:1999vm},
\begin{tcolorbox}[title=Expected optimal SNR valid for weak signal strength (earlier work)]
\begin{equation}
\label{eq:SNRold}
\textrm{SNR} \approx \left[T_{\rm obs} \int_{f_{\rm min}}^{f_{\rm max}}{\rm d}f \:\left(\frac{h^2\Omega_{\rm signal}(f)}{h^2\Omega_{\rm noise}(f)}\right)^2\right]^{1/2} \,,
\end{equation}
\end{tcolorbox}

\noindent
with the different quantities appearing in this expression being defined as follows: $T_{\rm obs}$ is the total accumulated observing time, the frequencies $f_{\rm min}$ and $f_{\rm max}$ span the LISA frequency band $[f_{\rm min},f_{\rm max}]$, and $h^2\Omega_{\rm signal}$ and $h^2\Omega_{\rm noise}$ describe the GWB signal strength and LISA noise power spectral density (PSD), respectively, both in units of $h^2\Omega_{\rm GW}$, i.e., the GW energy-density PSD (see Eq.~\eqref{eq:OGW} below for a more precise definition).

However, as emphasized in the title of the equation, the expression in Eq.~\eqref{eq:SNRold} is valid only in the limit of a weak GWB signal, $h^2\Omega_{\rm signal} \ll h^2\Omega_{\rm noise}$ at all relevant frequencies. This approximation is well justified as long as one is only interested in signals close to detection threshold, i.e., signals resulting in SNR values of $\mathcal{O}(1)$. Specifically, Equation~\eqref{eq:SNRold} is an appropriate starting point for the construction of standard LISA sensitivity curves, such as LISA's power-law-integrated (PLI) sensitivity  curve~\cite{Thrane:2013oya} or peak-integrated sensitivity curves~\cite{Schmitz:2020syl} for GWB signals from cosmological phase transitions. These sensitivity curves are commonly normalized to an SNR level of $1$, which justifies the use of Eq.~\eqref{eq:SNRold}.

At the same time, it is important to be aware of the assumptions that enter the derivation of Eq.~\eqref{eq:SNRold}. A crucial aspect in this regard is that, in the presence of a GWB signal, the full LISA data stream obviously receives two contributions: one from noise and one from the signal. However, when evaluating the noise variance in the weak-signal limit, one replaces the total auto-correlated PSD of the LISA data stream by the noise-only PSD, neglecting the signal contribution. This approximation can also be formulated in the language of hypothesis testing. The SNR of the auto-correlation statistic $S$ is defined as the ratio of $\mu$ and $\sigma$, where $\mu$ is the difference in the expectation values of $S$ under the signal and null hypotheses, $\mu = \langle S\rangle_{\mu \neq 0} - \langle S\rangle_{\mu = 0}$, and $\sigma$ is the positive square root of the variance  $\sigma^2 = \langle S^2\rangle-\langle S\rangle^2$. The SNR in the weak-signal limit can then be identified with%
\footnote{Here and throughout the rest of the paper, the subscripts $\mu \neq 0$ and $\mu=0$ are a shorthand notation for ``evaluated under the signal hypothesis'' and ``evaluated under the null hypothesis'', respectively.}
\begin{equation}
\label{eq:SNRweak}
\textrm{SNR} \approx \frac{\mu}{\left(\sigma^2\right)^{1/2}_{\mu = 0}} = \frac{\langle S\rangle_{\mu \neq 0} - \langle S\rangle_{\mu = 0}}{\left( \langle S^2\rangle-\langle S\rangle^2\right)^{1/2}_{\mu = 0}} \qquad \textrm{(weak signal strength)} \,,
\end{equation}
which is the ratio of $\mu$ under the signal hypothesis and $\sigma$ under the null hypothesis. Note that we assumed perfect knowledge of the detector noise in the definition of $\mu$. Equation~\eqref{eq:SNRweak} is appropriate for assessing the statistical significance of a detection claim.

However, for parameter estimation, the situation is different. If we already know that a GWB signal is present in the data and we want to describe its properties in more detail, we have to work under the signal hypothesis and include the signal contribution to the variance. Doing so will result in an SNR expression valid at arbitrary signal strength,
\begin{equation}
\textrm{SNR} =\frac{\mu}{\left(\sigma^2\right)^{1/2}_{\mu \neq 0}} = \frac{\langle S\rangle_{\mu \neq 0} - \langle S\rangle_{\mu = 0}}{\left( \langle S^2\rangle-\langle S\rangle^2\right)^{1/2}_{\mu \neq 0}} \qquad \textrm{(arbitrary signal strength)} \,.
\end{equation}
In this paper, we shall evaluate this expression and thus derive a general result for the expected SNR of a LISA auto-correlation measurement that can be used for weak, intermediate, and strong GWB signals. While we present our detailed calculation further below, let us quote our final result already here in the Introduction for the busy reader,
\begin{tcolorbox}[title=Expected optimal SNR valid for arbitrary signal strength (this work)]
\begin{equation}
\label{eq:SNRnew}
\textrm{SNR} = \left[T_{\rm obs} \int_{f_{\rm min}}^{f_{\rm max}}{\rm d}f \:\left(\frac{h^2\Omega_{\rm signal}(f)}{h^2\Omega_{\rm noise}(f)+h^2\Omega_{\rm signal}(f)}\right)^2\right]^{1/2} \,.
\end{equation}
\end{tcolorbox}

This expression is remarkably simple. In essence, it states that the total noise budget relevant for the computation of the expected SNR is the sum of the detector noise and the GWB signal. In other words, the GWB signal acts as another form of noise\,---\,GWB self-noise\,---\,that contributes to the variance of the detector data. The effect of GWB self-noise has been discussed before in the literature. For a recent example, see, e.g., Ref.~\cite{NANOGrav:2023ctt}, which discusses the impact of GWB self-noise on the sensitivity curve of the NANOGrav PTA. Meanwhile, a discussion of GWB self-noise in more general terms for a cross-correlation measurement in a detector network can be found in Ref.~\cite{Allen:1997ad}. Despite these resources in the literature, most authors, however, ignore the possible impact of GWB self-noise and work with SNR expressions along the lines of Eq.~\eqref{eq:SNRold}. A notable exception is Ref.~\cite{Saikawa:2018rcs}, which partially implements the formalism presented in Ref.~\cite{Allen:1997ad}.

The discussion in Ref.~\cite{Allen:1997ad} covers the case of a general cross-correlation measurement, but does not explicitly spell out how to apply the respective formulas to the special case of an auto-correlation measurement. The purpose of the present paper is to close this gap. With this goal in mind, the remainder of this article is organized as follows: in Sec.~\ref{sec:detector}, we will first review the construction of $h^2\Omega_{\rm noise}$, i.e., the LISA noise PSD in units of $h^2\Omega_{\rm GW}$, before we turn to the derivation of the expected SNR in the presence of an arbitrarily strong GWB signal in Sec~\ref{sec:snr}. Based on our master formula for the SNR, we will subsequently revise the construction of PLI sensitivity curves at high SNR levels in Sec.~\ref{sec:plots} and illustrate our results in terms of novel SNR contour plots (the numerical data behind these plots are available on Zenodo~\cite{zenodo}). As part of this discussion, we will also present an example of how to work with our SNR contour plots in the context of a specific GWB model: stable cosmic strings in the Nambu--Goto approximation~\cite{Wachter:2024aos,Wachter:2024zly}. Section~\ref{sec:conclusions} contains our conclusions. 


\section{LISA noise power spectral density}
\label{sec:detector}


\subsection{Power spectral densities}


In the presence of a GWB signal, the data stream $d_I$ of a GW detector $I$ receives a signal and a noise contribution,
\begin{equation}
d_I(t) = s_I(t) + n_I(t) \,.
\end{equation}
If both the signal and the noise are stochastic and Gaussian, their statistical properties are fully captured by their respective PSDs, i.e., the two-point correlators in Fourier space,
\begin{align}
\label{eq:sPSD}
\langle \tilde{s}_I(f)\,\tilde{s}_I^*(f')\rangle & = \frac{1}{2}\,\delta(f-f')\,\mathcal{R}_I(f)\,S_{\rm signal}(f) \,. \\
\label{eq:nPSD}
\langle\tilde{n}_I(f)\,\tilde{n}_I^*(f')\rangle & = \frac{1}{2}\,\delta(f-f')\,D_{\rm noise}^I(f) \,.
\end{align}
Here, $\tilde{s}_I$ and $\tilde{n}_I$ denote the Fourier transforms of $s_I$ and $n_I$, respectively; $\mathcal{R}_I$ is the detector transfer function or signal response function, which follows from convoluting the tensor perturbations describing the GWB signal with the detector's impulse response; $S_{\rm signal}$ is the strain signal PSD; and $D_{\rm noise}^I$ is the detector noise PSD. The factor of \sfrac{1}{2} accounts for the fact that we define all PSDs in this paper to be single-sided PSDs. For more details on these quantities and the underlying notation and conventions, see App.~A of Ref.~\cite{Schmitz:2020syl}. 

In order to facilitate the comparison between the properties of the signal and the noise, we can use $\mathcal{R}_I$ to convert the detector noise PSD $D_{\rm noise}^I$ to the strain noise PSD $S_{\rm noise}^I$,
\begin{equation}
S_{\rm noise}^I(f) = \frac{D_{\rm noise}^I(f)}{\mathcal{R}_I(f)} \,.
\end{equation}
Furthermore, the strain signal and noise PSDs can both be expressed in units of $h^2\Omega_{\rm GW}$, 
\begin{align}
\label{eq:Osignal}
h^2\Omega_{\rm signal}(f) & = \frac{2\pi^2}{3\,(H_0/h)^2}\,f^3 S_{\rm signal}(f) \,, \\
\label{eq:Onoise}
h^2\Omega_{\rm noise}^I(f) & = \frac{2\pi^2}{3\,(H_0/h)^2}\,f^3 S_{\rm noise}^I(f) \,,
\end{align}
where $h^2\Omega_{\rm GW}$ is the GW energy-density PSD, i.e., the GW energy density per logarithmic frequency interval in units of the present-day critical energy density of the Universe,
\begin{equation}
\label{eq:OGW}
h^2\Omega_{\rm GW} = \frac{1}{\rho_{\rm crit}/h^2} \,\frac{d\rho_{\rm GW}}{d\,\ln f} \,, \qquad \rho_{\rm crit}/h^2 = 3\,(H_0/h)^2 M_{\rm Pl}^2 \,.
\end{equation}
Here, $M_{\rm Pl} = (8\pi G)^{-1/2}\simeq 2.435 \times 10^{18}\,\textrm{GeV}$ denotes the reduced Planck mass, with $G$ being Newton's constant, and $H_0 = 100\,h\,\textrm{km}/\textrm{s}/\textrm{Mpc}$ is the Hubble constant, i.e., the present-day value of the Hubble rate, expressed in terms of the dimensionless Hubble constant $h$. The ratio $H_0/h$ is independent of the actual value of $h$ and can be regarded as a constant with units of a frequency, $H_0/h = 100\,\textrm{km}/\textrm{s}/\textrm{Mpc} \simeq 3.24 \times 10^{-18}\,\textrm{Hz}$. Hence, we work with $h^2\Omega_{\rm GW}$ instead of $\Omega_{\rm GW}$ in order to eliminate the dependence on the Hubble constant.


\subsection{Instrumental noise}


Let us now focus on the LISA noise PSD and discuss the various contributions to $S_{\rm noise}$. In doing so, we will drop the index $I$ as all relevant quantities will refer to LISA. In our discussion, we will follow closely Refs.~\cite{Schmitz:2020syl,Schmitz:2020rag}, where more details can be found. Specifically, following Ref.~\cite{Schmitz:2020rag}, we shall account for two independent contributions to $S_{\rm noise}$: LISA's instrumental strain noise, $S_{\rm inst}$, and confusion noise from galactic binaries, $S_{\rm gcn}$, 
\begin{equation}
\label{eq:Snoise}
S_{\rm noise}\left(f\right) = S_{\rm inst}\left(f\right) + S_{\rm gcn}\left(f\right) \,.
\end{equation}
Our main argument in this paper on the validity of the expression in Eq.~\eqref{eq:SNRold} hinges in no way on the inclusion of the confusion-noise term, though, and a generalization of our analysis to noise models including extra astrophysical foregrounds would be straightforward. 


The instrumental strain noise PSD can again be written as $S_{\rm inst} = D_{\rm inst} / \mathcal{R}$, where the detector noise PSD $D_{\rm inst}$ receives two stationary contributions~\cite{Robson:2018ifk} (see also Ref.~\cite{Edwards:2020tlp}),
\begin{equation}
\label{eq:DLISA}
D_{\rm inst}\left(f\right) = \frac{1}{L^2}\,D_{\rm oms}\left(f\right) + \frac{2}{\left(2\pi f\right)^4L^2}\left[1 + \cos^2\left(\frac{f}{f_*}\right)\right] D_{\rm acc}\left(f\right) \,.
\end{equation}
Here, $L = 2.5\times10^9\,\textrm{m}$ stands for the length of LISA's interferometer arms; $f_* = c/\left(2\pi L\right) \simeq 19.09\,\textrm{mHz}$ is LISA's transfer frequency~\cite{Robson:2018ifk} (closely related to the free-spectral-range frequency, $f_* = f_{\rm fsr}/\pi$); $D_{\rm oms}$ describes the noise in the optical metrology system (OMS) (i.e., position noise); and $D_{\rm acc}$ accounts for the acceleration noise of a single test mass,
\begin{align}
\label{eq:Doms}
D_{\rm oms}\left(f\right) & \simeq \left(1.5 \times 10^{-11}\,\textrm{m}\right)^2 \left[1 + \left(\frac{2\,\textrm{mHz}}{f}\right)^4\right] \textrm{Hz}^{-1} \,, \\
\label{eq:Dacc}
D_{\rm acc}\left(f\right) & \simeq \left(3\times10^{-15}\,\textrm{m}\,\textrm{s}^{-2}\right)^2 \left[1 + \left(\frac{0.4\,\textrm{mHz}}{f}\right)^2\right] \left[1 + \left(\frac{f}{8\,\textrm{mHz}}\right)^4\right] \textrm{Hz}^{-1} \,.
\end{align}


The LISA detector will consist of three spacecraft whose relative separations will be monitored by six inter-spacecraft laser links. By forming suitable time-delay interferometry (TDI) combinations of these measurements, one will be able to construct two approximately independent data streams ($A$ and $E$) that behave like equal-arm Michelson interferometers.%
\footnote{Describing LISA in terms of static equal-arm Michelson interferometers is an idealization. In reality, LISA will ultimately feature unequal arm lengths (see Refs.~\cite{Wang:2020fwa,Vallisneri:2020otf,Shi:2026kfk} for a more detailed discussion).}
A third TDI combination ($T$) will be strongly suppressed with respect to GW signals at low frequencies and can be used for instrumental noise characterization. The function $\mathcal{R}$ describes the combined response of LISA's $A$ and $E$ channels to an incoming GW~\cite{Larson:1999we,Liang:2019pry,Zhang:2019oet,Lu:2019log}. In the following, we shall work with the analytical expression provided in Ref.~\cite{Zhang:2020khm},
\begin{align}
\label{eq:R}
u^2\,\mathcal{R}\left(u,\gamma\right)
& = s_{2u}\left[s_{\gamma/2}^2\left(\frac{1}{u}+\frac{2}{u^3}\right) + c_{\gamma/2}^2\left(2\,\textrm{Si}\left(2u\right) - \textrm{Si}\left(2u_+\right) - \textrm{Si}\left(2u_-\right) \vphantom{\ln c_{\gamma/2}^2} \right)\right]
\\ \nonumber
& \, + c_{2u}\left[s_{\gamma/2}^2\left(\frac{1}{6}-\frac{2}{u^2}\right) + c_{\gamma/2}^2\left(2\,\textrm{Ci}\left(2u\right) - \textrm{Ci}\left(2u_+\right) - \textrm{Ci}\left(2u_-\right) + \ln c_{\gamma/2}^2\right)\right]
\\ \nonumber
& \, - \frac{s_{u_+-u_-}}{32u\,s_{\gamma/2}^3}\left(21-28\,c_\gamma+7\,c_{2\gamma} + \frac{3-c_\gamma}{u^2}\right) + \frac{c_{u_+-u_-}}{8u^2 s_{\gamma/2}^2}\left(1+s_{\gamma/2}^2 \vphantom{\frac{3}{u^2}} \right)
\\ \nonumber
& \, - 2\,s_{\gamma/2}^2\left(\textrm{Ci}\left(2u\right) - \textrm{Ci}\left(u_+-u_-\right) + \ln s_{\gamma/2}\right) + \frac{3-c_\gamma}{12} - \frac{1-c_\gamma}{u^2} \,,
\end{align}
where $s_x$, $c_x$, and $u_\pm$ are a shorthand notation for
\begin{align}
s_x = \sin\left(x\right) \,, \quad c_x = \cos\left(x\right) \,, \quad u_\pm = u \pm u\,\sin\left(\frac{\gamma}{2}\right) \,,
\end{align}
and $\textrm{Si}$ and $\textrm{Ci}$ denote the sine and cosine integral functions,
\begin{equation}
\textrm{Si}\left(x\right) = \int_0^x dt\:\frac{\sin\left(t\right)}{t} \,, \quad \textrm{Ci}\left(x\right) = -\int_x^\infty dt\:\frac{\cos\left(t\right)}{t} \,.
\end{equation}
The independent variables in Eq.~\eqref{eq:R} are $u = f/f_*$, i.e., the GW frequency in units of the transfer frequency, and the opening angle of the interferometers, $\gamma = \pi/3$, reflecting the equilateral triangular formation of the three LISA spacecraft.


\subsection{Galactic confusion noise}


In addition to instrumental strain noise, we shall account for galactic confusion noise (GCN) from galactic binaries. To this end, we will work with the semianalytical fit function in Ref.~\cite{Cornish:2017vip}, which is based on the compact-binary population model in Refs.~\cite{Nelemans:2004qz,Toonen:2012jj},
\begin{equation}
\label{eq:gcn}
S_{\rm gcn}\left(f\right) = A\,\left(\frac{1\,\textrm{mHz}}{f}\right)^{7/3} e^{-\left(f/f_{\rm ref}\right)^\alpha - \beta f\sin\left(\kappa f\right)}\left[1 + \tanh\left(\gamma\left(f_{\rm knee}-f\right)\right)\right] \,.
\end{equation}
Here, the overall amplitude $A$ and reference frequency $f_{\rm ref}$ are fixed at $A =  9 \times 10^{-38}\,\textrm{Hz}^{-1}$ and $f_{\rm ref} = 1000\,\textrm{mHz}$, respectively, while all other fit parameters depend on the observing time $T_{\rm obs}$ (see Tab.~\ref{tab:confusion}). The reason behind this time dependence is that, over the course of the LISA mission, more and more compact binaries will be resolved individually, such that they can be subtracted from the unresolved GNC background. In our analysis below, we will fix $T_{\rm obs} = 1\,\textrm{yr}$ and evaluate $S_{\rm gcn}$ accordingly based on the corresponding fit values in Tab.~\ref{tab:confusion}. A generalization of our analysis to other observing times would be straightforward.


\begin{table}
\begin{center} 
\renewcommand{\arraystretch}{1.22}
\caption{Fit parameters in the expression for $S_{\rm gcn}$ in Eq.~\eqref{eq:gcn} as functions of $T_{\rm obs}$~\cite{Cornish:2017vip}.}
\label{tab:confusion}
\medskip\smallskip
\begin{tabular}{|c||ccccc|}
\hline
$T_{\rm obs}$  $[\textrm{yr}]$       &
$\alpha$                             &
$\beta$        $[\textrm{mHz}^{-1}]$ &
$\kappa$       $[\textrm{mHz}^{-1}]$ &
$\gamma$       $[\textrm{mHz}^{-1}]$ &
$f_{\rm knee}$ $[\textrm{mHz}]$                                 \\ 
\hline\hline
0.5 & $0.133$ & $\phantom{-}0.243$ & $0.482$ & $0.917$ & $2.58$ \\
1.0 & $0.171$ & $\phantom{-}0.292$ & $1.020$ & $1.680$ & $2.15$ \\
2.0 & $0.165$ & $\phantom{-}0.299$ & $0.611$ & $1.340$ & $1.73$ \\
4.0 & $0.138$ & $          -0.221$ & $0.521$ & $1.680$ & $1.13$ \\
\hline
\end{tabular}
\end{center} 
\end{table}

\begin{figure}
\begin{center}
\includegraphics[width=0.67\textwidth]{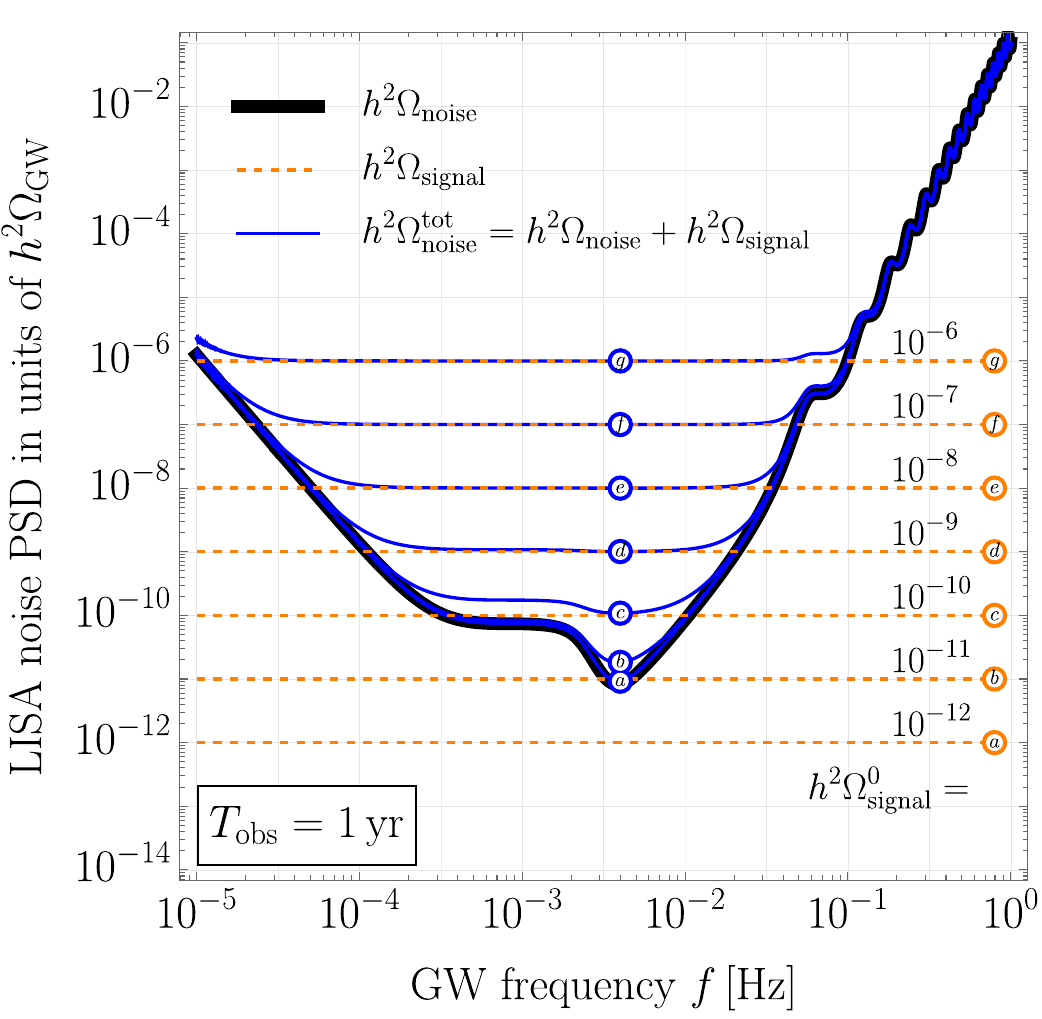}
\end{center}
\caption{LISA strain noise curve in units of $h^2\Omega_{\rm GW}$, with and without contributions from GWB self-noise. Without GWB self-noise, the LISA strain noise is given by the noise PSD at the detector level, $h^2\Omega_{\rm noise}$ (thick solid black line); with GWB self-noise, however, the total noise PSD $h^2\Omega_{\rm noise}^{\rm tot}$ (solid blue lines) also depends on the GWB signal $h^2\Omega_{\rm signal}$ itself (see Eq.~\eqref{eq:Onoisetot}). In this plot, we consider seven different possible signals: flat plateaus with amplitudes $h^2\Omega_{\rm signal}^0 = 10^{-(12\cdots 6)}$ (dashed orange lines). The markers (\textit{a}, \textit{b}, \textit{c}, etc.) identify pairs of $h^2\Omega_{\rm noise}^{\rm tot}$ and $h^2\Omega_{\rm signal}$ curves.}
\label{fig:noisePSD}
\end{figure}

Combining the expressions in Eqs.~\eqref{eq:Onoise}, \eqref{eq:Snoise}, \eqref{eq:DLISA}, \eqref{eq:Doms}, \eqref{eq:Dacc}, \eqref{eq:R}, \eqref{eq:gcn}, we are now able to compute the LISA noise PSD in units of $h^2\Omega_{\rm GW}$, i.e., $h^2\Omega_{\rm noise}$. Our central argument in this paper, however, is that the noise PSD $h^2\Omega_{\rm noise}$ is not sufficient to describe all forms of noise that are relevant for the computation of the expected SNR in the presence of a GWB signal\,---\,$h^2\Omega_{\rm noise}$ needs to be combined with the GWB signal itself,
\begin{equation}
\label{eq:Onoisetot}
h^2\Omega_{\rm noise}^{\rm tot}(f) = h^2\Omega_{\rm noise}(f) + h^2\Omega_{\rm signal}(f) \,,
\end{equation}
where the signal contribution can be interpreted as GWB self-noise (see our discussion in Sec.~\ref{sec:introduction}). In order to illustrate the effect of the GWB self-noise contribution, we plot $h^2\Omega_{\rm noise}$ next to $h^2\Omega_{\rm noise}^{\rm tot}$ for different choices of a scale-invariant signal, $h^2\Omega_{\rm signal}(f) = h^2\Omega_{\rm signal}^0$, in Fig.~\ref{fig:noisePSD}. As evident from this plot, the GWB self-noise can easily dominate the overall noise budget. For $h^2\Omega_{\rm  signal}^0 \gtrsim 10^{-9}$, e.g., the total noise PSD $h^2\Omega_{\rm noise}^{\rm tot}$ is well approximated by the signal contribution alone across the range of frequencies where LISA is most sensitive. 


\section{Expected signal-to-noise ratio}
\label{sec:snr}


\subsection{Auto-correlation statistic}


With the noise PSD $h^2\Omega_{\rm noise}$ at our disposal, we are now ready to turn to the computation of the expected SNR. First, let us define the relevant auto-correlation statistic,
\begin{equation}
\label{eq:S}
S = \int_{-T_{\rm obs}/2}^{T_{\rm obs}/2} {\rm d}t \int_{-T_{\rm obs}/2}^{T_{\rm obs}/2} {\rm d}t' \:  d(t)\,d(t')\, Q(t-t') \,.
\end{equation}
Then, introducing the expectation value $\mu$ and variance $\sigma^2$ of this statistic (see Sec.~\ref{sec:introduction}),
\begin{equation}
\mu = \langle S\rangle_{\mu \neq 0} - \langle S\rangle_{\mu = 0} \,, \qquad \sigma^2 = \langle S^2\rangle_{\mu \neq 0} - \langle S\rangle_{\mu \neq 0}^2 \,,  
\end{equation}
we can define our main quantity of interest: the expectation value of the optimal SNR,
\begin{equation}
\textrm{SNR} = \frac{\mu}{\sigma} = \frac{\langle S\rangle_{\mu \neq 0} - \langle S\rangle_{\mu = 0}}{\left( \langle S^2\rangle-\langle S\rangle^2\right)^{1/2}_{\mu \neq 0}} \,.
\end{equation}
This expected SNR is optimal in the sense that the filter function $Q$ in Eq.~\eqref{eq:S} must be matched to the signal present in the data so as to maximize the ratio $\mu/\sigma$, i.e., $Q$ is a matched filter. We shall now compute $\mu$ and $\sigma$ and determine the correct choice of filter.


\subsection{Derivation for arbitrary signal strength}


In order to evaluate the expectation values in the expressions for $\mu$ and $\sigma$, it is convenient to rewrite $S$ in terms of the Fourier transforms of the data stream and the filter~\cite{Allen:1997ad},
\begin{equation}
\label{eq:SFourier}
S = \int_{-\infty}^{\infty} {\rm d} f \int_{-\infty}^{\infty} {\rm d} f'\: \delta_{T_{\rm obs}}(f-f')\, \tilde{d}(f)\,\tilde{d}^*(f')\,\tilde{Q}(f) \,,
\end{equation}
where $\delta_{T_{\rm obs}}$ denotes the finite-time approximation to the Dirac delta function,
\begin{equation}
\delta_{T_{\rm obs}}(f-f') = T_{\rm obs}\:\textrm{sinc}(\pi fT_{\rm obs}) = T_{\rm obs}\:\frac{\sin(\pi fT_{\rm obs})}{\pi fT_{\rm obs}} \,. 
\end{equation}
Given the expression in Eq.~\eqref{eq:SFourier}, it is now easy to write down the expectation value of $S$,
\begin{equation}
\label{eq:Sdd}
\langle S\rangle_{\mu \neq 0} = \int_{-\infty}^{\infty} {\rm d} f \int_{-\infty}^{\infty} {\rm d} f'\: \delta_{T_{\rm obs}}(f-f')\, \langle\tilde{d}(f)\,\tilde{d}^*(f')\rangle \,\tilde{Q}(f) \,.
\end{equation}
Since the GWB signal and the noise are uncorrelated, Equations~\eqref{eq:sPSD} and \eqref{eq:nPSD} imply
\begin{equation}
\langle S\rangle_{\mu \neq 0} = \frac{1}{2}\int_{-\infty}^{\infty} {\rm d} f \int_{-\infty}^{\infty} {\rm d} f'\: \delta_{T_{\rm obs}}(f-f')\,\delta(f-f')\,\mathcal{R}(f)\left[S_{\rm signal}(f)+S_{\rm noise}(f)\right]\tilde{Q}(f) \,.
\end{equation}
Using the Dirac delta function to eliminate the integral over $f'$, this can be simplified to
\begin{equation}
\label{eq:Sexp}
\langle S\rangle_{\mu \neq 0} = \frac{T_{\rm obs}}{2}\int_{-\infty}^{\infty} {\rm d} f \: \mathcal{R}(f)\left[S_{\rm signal}(f)+S_{\rm noise}(f)\right]\tilde{Q}(f) \,,
\end{equation}
where we inserted $\delta_{T_{\rm obs}}(0) = T_{\rm obs}$. The expression in Eq.~\eqref{eq:Sexp} represents the expectation value of $S$ under the signal hypothesis. Following Ref.~\cite{Thrane:2013oya}, we shall now assume that we have perfect knowledge of the detector noise (e.g., thanks to measurements in the $T$ channel), such that we exactly know the expectation value of $S$ under the null hypothesis, 
\begin{equation}
\langle S\rangle_{\mu = 0} = \frac{T_{\rm obs}}{2}\int_{-\infty}^{\infty} {\rm d} f \: \mathcal{R}(f)\,S_{\rm noise}(f)\,\tilde{Q}(f) \,.
\end{equation}
The mean, $\mu$, is defined as the difference between the expectation value and this noise floor,
\begin{equation}
\label{eq:mu}
\mu = \langle S\rangle_{\mu \neq 0} - \langle S\rangle_{\mu = 0} = \frac{T_{\rm obs}}{2}\int_{-\infty}^{\infty} {\rm d} f \: \mathcal{R}(f)\,S_{\rm signal}(f)\,\tilde{Q}(f) \,.
\end{equation}

Next, we turn to the variance of the auto-correlation statistic. Note that, for the discussion of this quantity, the subtraction of the noise contribution from the total expectation value in Eq.~\eqref{eq:mu} is of no relevance. To see this, consider the shifted statistic 
\begin{equation}
S' = S - \langle S\rangle_{\mu = 0} \,.
\end{equation}
It is then easy to show that the variance of $S'$ is identical to the variance of $S$,
\begin{equation}
\sigma_{S'}^2 = \langle S'^2 \rangle_{\mu \neq 0} - \langle S' \rangle_{\mu \neq 0}^2 \,, \qquad \sigma_S^2 = \langle S^2 \rangle_{\mu \neq 0} - \langle S \rangle_{\mu \neq 0}^2 \qquad\Rightarrow\qquad \sigma_{S'}^2 = \sigma_{S}^2 \,.
\end{equation}

In a first step towards $\sigma^2$, we return to $\langle S\rangle_{\mu \neq 0}$ in Eq.~\eqref{eq:Sdd} and take its square,
\begin{align}
\label{eq:Var1}
\langle S\rangle_{\mu \neq 0}^2 = \int_{-\infty}^{\infty} {\rm d} f \int_{-\infty}^{\infty} {\rm d} f' \int_{-\infty}^{\infty} {\rm d} f'' \int_{-\infty}^{\infty} {\rm d} f''' \: \delta_{T_{\rm obs}}(f-f')\,\delta_{T_{\rm obs}}(f''-f''') \\ \nonumber
\times\, \langle\tilde{d}(f)\,\tilde{d}^*(f')\rangle\, \langle\tilde{d}(f'')\,\tilde{d}^*(f''')\rangle \,\tilde{Q}(f) \, \tilde{Q}(f'') \,,
\end{align}
which needs to be subtracted from the expectation value of $S^2$ under the signal hypothesis,
\begin{align}
\label{eq:Var2}
\langle S^2\rangle_{\mu \neq 0} = \int_{-\infty}^{\infty} {\rm d} f \int_{-\infty}^{\infty} {\rm d} f' \int_{-\infty}^{\infty} {\rm d} f'' \int_{-\infty}^{\infty} {\rm d} f''' \: \delta_{T_{\rm obs}}(f-f')\,\delta_{T_{\rm obs}}(f''-f''') \\ \nonumber
\times\, \langle\tilde{d}(f)\,\tilde{d}^*(f')\,\tilde{d}(f'')\,\tilde{d}^*(f''')\rangle \,\tilde{Q}(f) \, \tilde{Q}(f'') \,.
\end{align}
The difference between these two expressions is only in the integrand, which contains a product of two two-point correlators in the case of $\langle S\rangle_{\mu \neq 0}^2$ and simply one four-point correlator in the case of $\langle S^2\rangle_{\mu \neq 0}$. Since we are only interested in Gaussian random variables (both at the signal and the noise level), we can use Isserlis's theorem in probability theory (also known as Wick's probability theorem) to write the four-point correlator as
\begin{align}
& \langle\tilde{d}(f)\,\tilde{d}^*(f')\,\tilde{d}(f'')\,\tilde{d}^*(f''')\rangle \\ \nonumber
&\hspace{0.25in} = \langle\tilde{d}(f)\,\tilde{d}^*(f')\rangle\, \langle\tilde{d}(f'')\,\tilde{d}^*(f''')\rangle \\ \nonumber
&\hspace{0.25in} + \langle\tilde{d}(f)\,\tilde{d}(f'')\rangle\, \langle\tilde{d}^*(f')\,\tilde{d}^*(f''')\rangle  \\ \nonumber
&\hspace{0.25in} + \langle\tilde{d}(f)\,\tilde{d}^*(f''')\rangle\, \langle\tilde{d}^*(f')\,\tilde{d}(f'')\rangle \,.
\end{align}

Subtracting Eq.~\eqref{eq:Var1} from Eq.~\eqref{eq:Var2} thus allows us to write $\sigma^2 = \sigma_1^2 + \sigma_2^2$, where
\begin{align}
\sigma_1^2 = \int_{-\infty}^{\infty} {\rm d} f \int_{-\infty}^{\infty} {\rm d} f' \int_{-\infty}^{\infty} {\rm d} f'' \int_{-\infty}^{\infty} {\rm d} f''' \: \delta_{T_{\rm obs}}(f-f')\,\delta_{T_{\rm obs}}(f''-f''') \\ \nonumber
\times\, \langle\tilde{d}(f)\,\tilde{d}(f'')\rangle\, \langle\tilde{d}^*(f')\,\tilde{d}^*(f''')\rangle  \,\tilde{Q}(f) \, \tilde{Q}(f'') \,, \\
\sigma_2^2 = \int_{-\infty}^{\infty} {\rm d} f \int_{-\infty}^{\infty} {\rm d} f' \int_{-\infty}^{\infty} {\rm d} f'' \int_{-\infty}^{\infty} {\rm d} f''' \: \delta_{T_{\rm obs}}(f-f')\,\delta_{T_{\rm obs}}(f''-f''') \\ \nonumber
\times\, \langle\tilde{d}(f)\,\tilde{d}^*(f''')\rangle\, \langle\tilde{d}^*(f')\,\tilde{d}(f'')\rangle  \,\tilde{Q}(f) \, \tilde{Q}(f'') \,.
\end{align}
In order to evaluate correlators of the form $\langle \tilde{d}\tilde{d}\rangle$ or $\langle \tilde{d}^*\tilde{d}^*\rangle$, we note that the data stream $d$ is real-valued, $d=d^*$, which implies $\tilde{d}^*(f) = \tilde{d}(-f)$ in Fourier space. We hence obtain
\begin{align}
\sigma_1^2 &= \frac{1}{4}\int_{-\infty}^{\infty} {\rm d} f \int_{-\infty}^{\infty} {\rm d} f' \: \delta_{T_{\rm obs}}(f-f')\,\delta_{T_{\rm obs}}(f-f') \\ \nonumber
&\hspace{0.2in}\times\, \mathcal{R}(f)\mathcal{R}(f')\left[S_{\rm signal}(f)+S_{\rm noise}(f)\right]\left[S_{\rm signal}(f')+S_{\rm noise}(f')\right] \,\tilde{Q}(f) \, \tilde{Q}(-f) \,, \\
\sigma_2^2 &= \frac{1}{4}\int_{-\infty}^{\infty} {\rm d} f \int_{-\infty}^{\infty} {\rm d} f' \: \delta_{T_{\rm obs}}(f-f')\,\delta_{T_{\rm obs}}(f-f') \\ \nonumber
&\hspace{0.2in}\times\, \mathcal{R}(f)\mathcal{R}(f')\left[S_{\rm signal}(f)+S_{\rm noise}(f)\right]\left[S_{\rm signal}(f')+S_{\rm noise}(f')\right] \,\tilde{Q}(f) \, \tilde{Q}(f') \,.
\end{align}
By construction, the filter function $Q$ is symmetric in its argument in the time domain, $Q(t-t')=Q(t'-t)$ (see Eq.~\eqref{eq:S}), which means that we can replace $\tilde{Q}(-f)=\tilde{Q}(f)$. Moreover, for large $T_{\rm obs}$, we may approximate $\delta_{T_{\rm obs}}^2(f-f') \approx \delta_{T_{\rm obs}}(f-f')\,\delta(f-f')$, so
\begin{equation}
\sigma_1^2 = \sigma_2^2 =  \frac{T_{\rm obs}}{4}\int_{-\infty}^{\infty} {\rm d} f \: \mathcal{R}^2(f)\left[S_{\rm signal}(f)+S_{\rm noise}(f)\right]^2 \tilde{Q}^2(f) \,.
\end{equation}
We thus find for the variance of the auto-correlation statistic under the signal hypothesis 
\begin{equation}
\label{eq:sigma2}
\sigma^2 =  \frac{T_{\rm obs}}{2}\int_{-\infty}^{\infty} {\rm d} f \: \mathcal{R}^2(f)\left[S_{\rm signal}(f)+S_{\rm noise}(f)\right]^2 \tilde{Q}^2(f) \,.
\end{equation}

Our argument in this paper builds on the observation that the result in Eq.~\eqref{eq:sigma2} contains both a noise and a signal (i.e., GWB self-noise) contribution. If we were to derive the standard expression for the SNR in the weak-signal limit in Eq.~\eqref{eq:SNRold}, we would have to drop the GWB self-noise contribution in Eq.~\eqref{eq:sigma2} in the following. The key difference between our analysis and earlier work, however, is that we will not do so, but instead keep the GWB self-noise contribution to the variance throughout the whole computation. 

To continue, let us first make the notation more compact and introduce the PSD $P$ for the total noise-plus-signal power seen by LISA. Since the noise and signal are uncorrelated,
\begin{equation}
\langle\tilde{d}(f)\,\tilde{d}^*(f')\rangle = \frac{1}{2}\,\delta(f-f')\,P(f) \,, \qquad P(f) = \mathcal{R}(f)\left[S_{\rm noise}(f)+S_{\rm signal}(f)\right] \,.
\end{equation}
In terms of the total-power PSD $P$, the variance $\sigma^2$ can be conveniently be written as
\begin{equation}
\sigma^2 = \frac{T_{\rm obs}}{2}\int_{-\infty}^{\infty} {\rm d} f \: P^2(f)\,\tilde{Q}^2(f) \,.
\end{equation}
To the best of our knowledge, this result for the variance of an auto-correlation search for a GWB signal, including the detector and GWB self-noise, has not appeared in the literature before. The corresponding expression for a cross-correlation search with a detector network can be found in Sec.\ V.A of Ref.~\cite{Allen:1997ad} (see Eqs.~(5.3) and (5.4) therein). Following  Ref.~\cite{Allen:1997ad}, we now introduce a positive-definite inner product for functions in the frequency domain,
\begin{equation}
(A,B) = \int_{-\infty}^{\infty} {\rm d} f \: A^*(f)\, B(f)\,P^2(f) \,,
\end{equation}
which allows us to summarize our results for $\mu$ and $\sigma$ in the following elegant fashion,
\begin{equation}
\mu = \frac{T_{\rm obs}}{2} \left(\mathcal{R}S_{\rm signal}/P^2,\tilde{Q}\right) \,, \qquad \sigma^2 =  \frac{T_{\rm obs}}{2} \left(\tilde{Q},\tilde{Q}\right) \,.
\end{equation}
Correspondingly, given these expressions for $\mu$ and $\sigma$, the expected SNR is given by
\begin{equation}
\label{eq:SNRinnerproducts}
\textrm{SNR} = \sqrt{\frac{T_{\rm obs}}{2}} \left(\mathcal{R}S_{\rm signal}/P^2,\tilde{Q}\right)\left(\tilde{Q},\tilde{Q}\right)^{-1/2} \,.
\end{equation}

In view of Eq.~\eqref{eq:SNRinnerproducts}, it is now easy to determine the optimal filter that maximizes the expected SNR. The Cauchy--Schwarz inequality for positive-definite inner products tells us that the scalar product between $\tilde{Q}$ and $\mathcal{R}S_{\rm signal}/P^2$ is largest when both functions point into the same direction in function space, i.e., if they are proportional to each other,
\begin{equation}
\label{eq:Q}
\tilde{Q}(f) \propto \mathcal{R}S_{\rm signal}/P^2 \,.
\end{equation}
This is all we need to know; the normalization of the filter is irrelevant because we explicitly divide by the norm of $\tilde{Q}$ in Eq.~\eqref{eq:SNRinnerproducts}. Substituting Eq.~\eqref{eq:Q} into Eq.~\eqref{eq:SNRinnerproducts} then yields
\begin{align}
\textrm{SNR} & = \sqrt{\frac{T_{\rm obs}}{2}} \left(\mathcal{R}S_{\rm signal}/P^2,\mathcal{R}S_{\rm signal}/P^2\right)^{1/2} \\
& = \left[\frac{T_{\rm obs}}{2}\int_{-\infty}^{\infty} {\rm d} f \: \left(\frac{\mathcal{R}(f)\,S_{\rm signal}(f)}{P(f)}\right)^2 \right]^{1/2} \\
& = \left[\frac{T_{\rm obs}}{2}\int_{-\infty}^{\infty} {\rm d} f \: \left(\frac{S_{\rm signal}(f)}{S_{\rm noise}(f)+S_{\rm signal}(f)}\right)^2 \right]^{1/2} \\
& = \left[T_{\rm obs}\int_0^{\infty} {\rm d} f \: \left(\frac{h^2\Omega_{\rm signal}(f)}{h^2\Omega_{\rm noise}(f)+h^2\Omega_{\rm signal}(f)}\right)^2 \right]^{1/2} \,,
\end{align}
where, in the last step, we restricted the integration domain to positive frequencies and replaced the strain signal and noise PSDs by their respective expressions in units of $h^2\Omega_{\rm GW}$. The noise PSD only has support in a finite frequency band $[f_{\rm min},f_{\rm max}]$; outside this band, we take $h^2\Omega_{\rm noise} \rightarrow \infty$, which leads us to our final expression for the expected SNR, 
\begin{equation}
\label{eq:SNRfinal}
\textrm{SNR} = \left[T_{\rm obs} \int_{f_{\rm min}}^{f_{\rm max}}{\rm d}f \:\left(\frac{h^2\Omega_{\rm signal}(f)}{h^2\Omega_{\rm noise}(f)+h^2\Omega_{\rm signal}(f)}\right)^2\right]^{1/2} \,.
\end{equation}


\begin{figure}
\begin{center}
\includegraphics[width=0.67\textwidth]{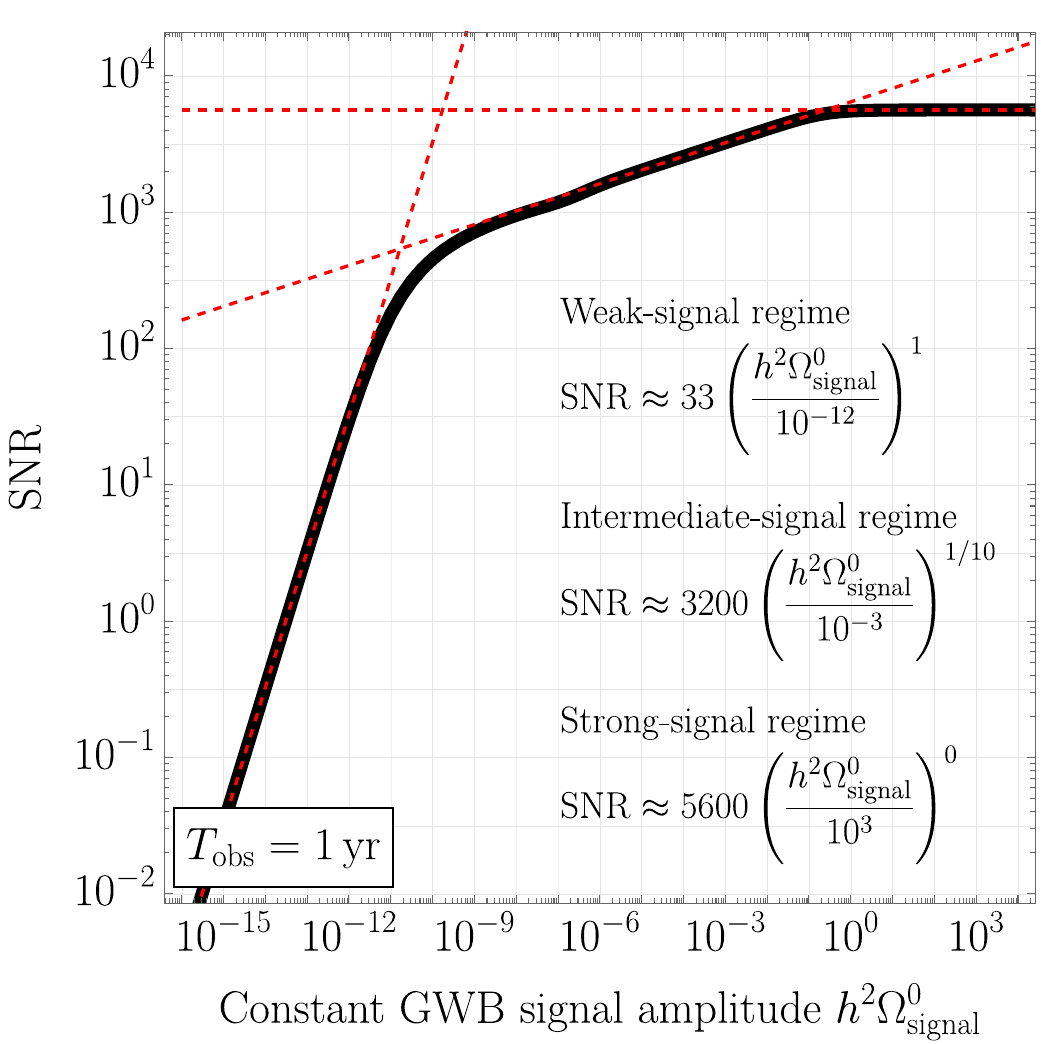}
\caption{Expected SNR for a flat GWB spectrum, $h^2\Omega_{\rm signal}(f) = h^2\Omega_{\rm signal}^0$, based on Eq.~\eqref{eq:SNRfinal} (solid black line), alongside the fit functions in Eqs.~\eqref{eq:SNRWSR}, \eqref{eq:SNRISR}, and \eqref{eq:SNRSSR} (dashed red lines).}
\label{fig:snrlogA}
\end{center}
\end{figure}


\subsection{Weak/intermediate/strong-signal regime}


Equation~\eqref{eq:SNRfinal} is a central result of this paper. The corresponding result for a LISA auto-correlation search in the weak-signal limit (see Eq.~\eqref{eq:SNRold}) can be found, e.g., in Eq.~(36) of Ref.~\cite{Thrane:2013oya}. Our result now generalizes this expression to arbitrary signal strength. In the next section, we will use Eq.~\eqref{eq:SNRfinal} to revise the construction of PLI sensitivity curves. However, before we do so, let us first investigate the behavior of the expected SNR as a function of increasing GWB signal strength. To this end, we will set, here and throughout the rest of the paper, $T_{\rm obs} = 1\,\textrm{yr}$ and $[f_{\rm min},f_{\rm max}] = [10^{-5}\,\,\textrm{Hz},1\,\,\textrm{Hz}]$, for definiteness.

In Fig.~\ref{fig:snrlogA}, we plot the expected SNR as a function of the amplitude of a constant, scale-invariant GWB signal, i.e., for GWB signals as indicated by the dashed orange lines in Fig.~\ref{fig:noisePSD}. As evident from Fig.~\ref{fig:snrlogA}, we need to distinguish between three qualitatively different regimes: the weak-signal regime (WSR), intermediate-signal regime (ISR), and strong-signal regime (SSR) (see Ref.~\cite{Siemens:2013zla} for a related discussion for PTAs).  In the weak-signal regime, GWB self-noise is negligible, and the SNR expression commonly employed in the literature (see Eq.~\eqref{eq:SNRold}) is applicable. In particular, for a scale-invariant GWB spectrum, the constant GWB amplitude $\,h^2\Omega_{\rm signal}^0$ can be factored out of the integral, 
\begin{equation}
\textrm{SNR} \:\overset{\textrm{(WSR)}}{\approx}\: \left[T_{\rm obs} \int_{f_{\rm min}}^{f_{\rm max}}{\rm d}f \:\left(\frac{1}{h^2\Omega_{\rm noise}(f)}\right)^2\right]^{1/2}\,h^2\Omega_{\rm signal}^0 \,.
\end{equation}
The prefactor can be easily computed numerically. For our choice of $T_{\rm obs}$ and $[f_{\rm min},f_{\rm max}]$, 
\begin{equation}
\left[T_{\rm obs} \int_{f_{\rm min}}^{f_{\rm max}}{\rm d}f \:\left(\frac{1}{h^2\Omega_{\rm noise}(f)}\right)^2\right]^{1/2} \simeq 3.3 \times 10^{13} \,,
\end{equation}
which allows us to write down a simple linear relation in the weak-signal regime,
\begin{equation}
\label{eq:SNRWSR}
\textrm{SNR} \:\overset{\textrm{(WSR)}}{\approx}\: 33 \left(\frac{h^2\Omega_{\rm signal}^0}{10^{-12}}\right)^1 \,.
\end{equation}
This expression reproduces the exact result with excellent precision for sufficiently weak signals. Around $h^2\Omega_{\rm signal}^0 \simeq 10^{-11}$, however, it begins to become inaccurate. This is an important observation, notably in view of the large number of analyses in the literature that work with the SNR in Eq.~\eqref{eq:SNRold} and GWB signals much stronger than $h^2\Omega_{\rm signal}^0 \simeq 10^{-11}$. 

At values of the GWB amplitude between $h^2\Omega_{\rm signal}^0 \simeq 10^{-11}$ and $h^2\Omega_{\rm signal}^0 \simeq 10^{-9}$, the expected SNR transitions from the weak-signal to the intermediate-signal regime. Deep inside the intermediate-signal regime, the expected SNR follows again a power law, 
\begin{equation}
\label{eq:SNRISR}
\textrm{SNR} \:\overset{\textrm{(ISR)}}{\approx}\: 3200 \left(\frac{h^2\Omega_{\rm signal}^0}{10^{-3}}\right)^{1/10} \,.
\end{equation}
This scaling can be easily understood by noting that the LISA strain noise curve in Fig.~\ref{fig:noisePSD} grows like $f^5$ at large frequencies (up to the sinusoidal modulations caused by $\mathcal{R}$), 
\begin{equation}
h^2\Omega_{\rm noise}(f) \approx 0.12 \left(\frac{f}{1\,\textrm{Hz}}\right)^5 \,, \qquad f \gtrsim 0.1\,\textrm{Hz} \,.
\end{equation}
As a rough approximation, we can thus model the integrand in the SNR integral as follows, 
\begin{equation}
\left(\frac{h^2\Omega_{\rm signal}(f)}{h^2\Omega_{\rm noise}(f)+h^2\Omega_{\rm signal}(f)}\right)^2 \:\overset{\textrm{(ISR)}}{\approx}\: \Theta\left(f_0-f\right) \,,
\end{equation}
with $f_0$ being the frequency where the constant GWB signal intersects the noise curve,
\begin{equation}
0.12 \left(\frac{f_0}{1\,\textrm{Hz}}\right)^5 = h^2\Omega_{\rm signal}^0 \qquad\Rightarrow\qquad f_0 = 1\,\textrm{Hz}\left(\frac{h^2\Omega_{\rm signal}^0}{0.12}\right)^{1/5} \,,
\end{equation}
and where $\Theta$ is the Heaviside theta function. With these approximations, we obtain 
\begin{equation}
\label{eq:SNRISR2}
\textrm{SNR} \:\overset{\textrm{(ISR)}}{\approx}\: \sqrt{T_{\rm obs}\left(f_0-f_{\rm min}\right)} \approx \sqrt{T_{\rm obs} f_0} \simeq 3500 \left(\frac{h^2\Omega_{\rm signal}^0}{10^{-3}}\right)^{1/10} \,,
\end{equation}
which reproduces the numerical fit function in Eq.~\eqref{eq:SNRISR} up to an $\mathcal{O}(10\,\%)$ deviation. In Fig.~\ref{fig:snrlogA}, we show the numerical fit in Eq.~\eqref{eq:SNRISR} and not the rough estimate in Eq.~\eqref{eq:SNRISR2}.

Finally, at GWB amplitudes around $h^2\Omega_{\rm signal}^0 \simeq 10^{-1}$, the frequency $f_0$ approaches $f_{\rm max}$, which marks the point where the intermediate-signal regime transitions to the strong-signal regime. In this regime, the expected SNR approaches a constant maximal value,
\begin{equation}
\label{eq:SNRSSR}
\textrm{SNR} \:\overset{\textrm{(SSR)}}{\approx}\: \textrm{SNR}_{\rm max} = \sqrt{T_{\rm obs}(f_{\rm max}-f_{\rm min})} \simeq 5600 \,,
\end{equation}
for our choice of $T_{\rm obs}$ and $[f_{\rm min},f_{\rm max}]$. In summary, we conclude that the ISR and SSR fit functions in Eqs.~\eqref{eq:SNRISR} and \eqref{eq:SNRSSR} deviate significantly from the standard WSR result in Eq.~\eqref{eq:SNRWSR}: for GWB amplitudes larger than  $h^2\Omega_{\rm signal}^0 \simeq 10^{-11}$, the growth of the expected SNR with the signal amplitude slows down dramatically, switching from a linear $(h^2\Omega_{\rm signal}^0)^1$ power law to a $(h^2\Omega_{\rm signal}^0)^{1/10}$ power law, and for GWB amplitudes larger than  $h^2\Omega_{\rm signal}^0 \simeq 10^{-1}$, the expected SNR is more or less saturated at a maximal value, $\textrm{SNR}_{\rm max}$.%
\footnote{Of course, GWB amplitudes as large as $h^2\Omega_{\rm signal}^0 \simeq 10^{-1}$ are not expected from a physical point of view. Our interest in this section pertains to the behavior of the expected SNR from a conceptual perspective.} 


\section{Signal-to-noise ratio contours}
\label{sec:plots}


\subsection{Power-law-integrated sensitivity curves}


\begin{figure}
\begin{center}
\includegraphics[width=0.67\textwidth]{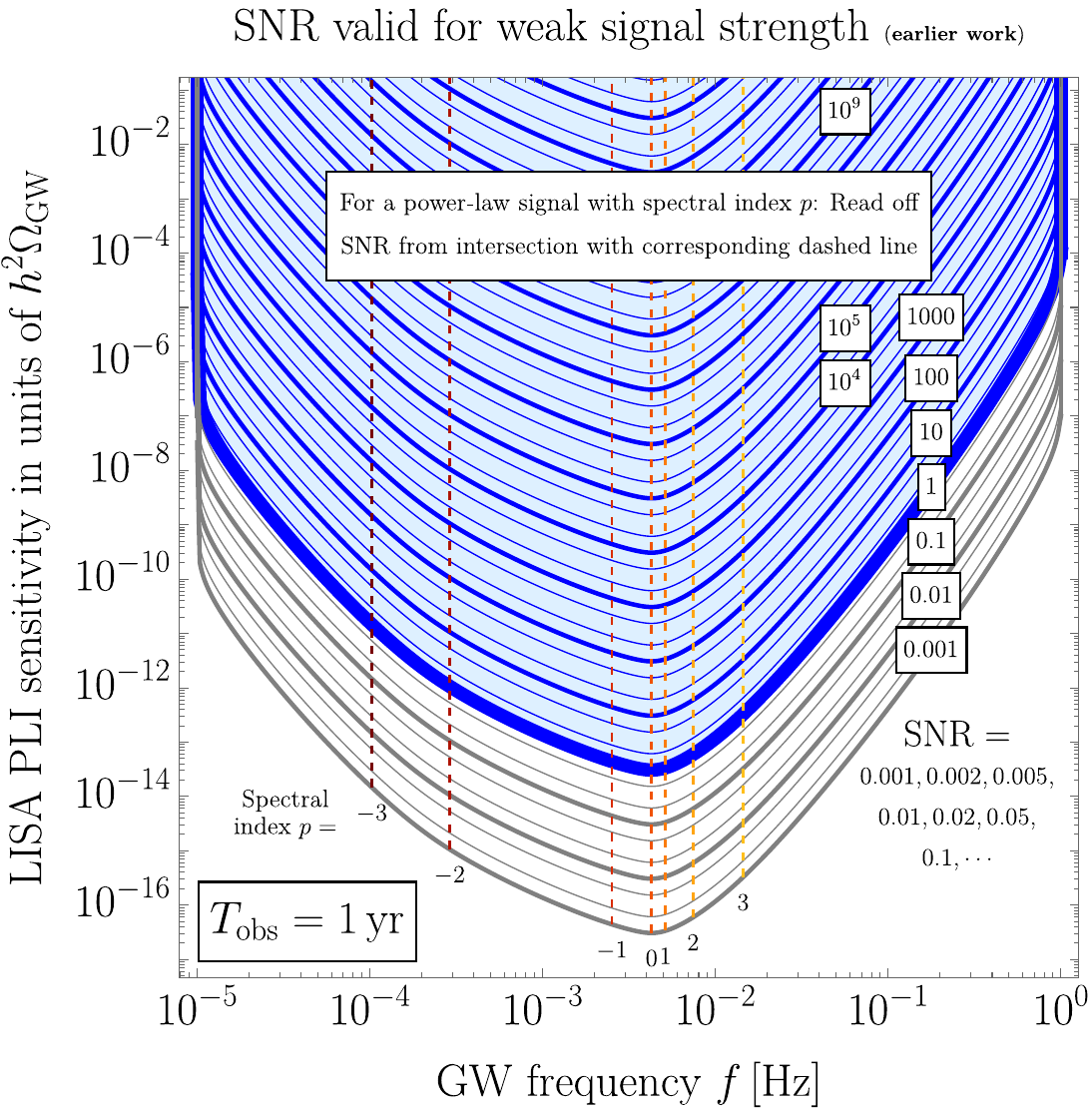}
\end{center}
\caption{PLI sensitivity curves derived from the SNR in the weak-signal regime (see Eq.~\eqref{eq:PLISWSR}).}
\label{fig:PLISold}
\end{figure}


The expected SNR is the basis for the construction of PLI sensitivity curves~\cite{Thrane:2013oya}, which allow for a graphical evaluation of the SNR in plots of the GW energy-density PSD. In this section, we shall generalize the existing construction of PLI sensitivity curves to GWB signals of arbitrary strength based on our master formula in Eq.~\eqref{eq:SNRnew}. However, before we do so, let us first review the construction in the weak-signal limit based on Eq.~\eqref{eq:SNRold}. 

Consider the SNR in Eq.~\eqref{eq:SNRold} in combination with a power-law GWB signal,
\begin{equation}
\textrm{SNR} \:\overset{\textrm{(WSR)}}{\approx}\: \left[T_{\rm obs} \int_{f_{\rm min}}^{f_{\rm max}}{\rm d}f \:\left(\frac{h^2\Omega_{\rm signal}(f)}{h^2\Omega_{\rm noise}(f)}\right)^2\right]^{1/2} \,, \qquad h^2\Omega_{\rm signal}(f) = A_p\left(\frac{f}{f_{\rm ref}}\right)^p \,,
\end{equation}
where $f_{\rm ref}$ is a conveniently chosen, but otherwise physically irrelevant, reference frequency. For every power-law index $p$, it is then straightforward to determine the required amplitude, $A_p$, that is necessary to achieve a given SNR level after a total observing time $T_{\rm obs}$,
\begin{equation}
\label{eq:ApWSR}
A_p^{\rm WSR}(\textrm{SNR}) = \textrm{SNR} \left[T_{\rm obs} \int_{f_{\rm min}}^{f_{\rm max}}{\rm d}f \:\left(\frac{(f/f_{\rm ref})^p}{h^2\Omega_{\rm noise}(f)}\right)^2\right]^{-1/2} \,.
\end{equation}
For fixed SNR, $T_{\rm obs}$, and $[f_{\rm min},f_{\rm max}]$, these amplitudes yield a family of power-law signals, the envelope of which defines the PLI sensitivity (PLIS) curve at the respective SNR level,
\begin{equation}
\label{eq:PLISWSR}
h^2\Omega_{\rm PLIS}^{\rm WSR}(f;\textrm{SNR}) = \max_p \left[A_p^{\rm WSR}(\textrm{SNR})\left(\frac{f}{f_{\rm ref}}\right)^p\right] \,.
\end{equation}

PLI sensitivity curves of this type are well established in literature. In Fig.~\ref{fig:PLISold}, we plot a large number of such standard PLI sensitivity curves for SNR levels ranging from $\textrm{SNR}=0.001$ to $\textrm{SNR}=10^9$ and even beyond. The interpretation of these curves is straightforward~\cite{Thrane:2013oya}: given a power-law GWB signal, one needs to find the PLI sensitivity curve to which the signal curve is tangent\,---\,the SNR level of the PLI sensitivity curve identified in this way then tells us the expected SNR for the GWB signal in question. To facilitate this graphical evaluation of the expected SNR, we also indicate in Fig.~\ref{fig:PLISold} where power-law GWB signals with specific values of the spectral index, $p=-3,-2,\cdots,2,3$, will actually be tangent to the family of PLI sensitivity curves; see the vertical colored dashed lines in Fig.~\ref{fig:PLISold}. In combination with these dashed lines, it is even easier to evaluate the expected SNR: for a given spectral index, $p=-3,-2,\cdots,2,3$, one simply needs to find the intersection of the signal curve with the respective dashed line\,---\,the SNR level of the PLI sensitivity curve going through this intersection point (to which the signal curve will automatically be tangent) then tells us the expected SNR for the GWB signal in question.

The PLI sensitivity curves in Fig.~\ref{fig:PLISold} are all parallel to each other, which is, of course, no coincidence. The regular appearance of these curves simply reflects the linear relation between the amplitudes $A_p$ in Eq.~\eqref{eq:ApWSR} and the selected SNR level. As a consequence, the information in the family of PLI sensitivity curves in Fig.~\ref{fig:PLISold} is redundant. It suffices to simply show one PLI sensitivity curve, conventionally the one for $\textrm{SNR} = 1$, which corresponds to what is typically done in the literature.  In this case, the vertical distance between the PLI sensitivity curve to which the signal curve is tangent (but which is maybe not explicitly plotted) and the reference PLI sensitivity curve for $\textrm{SNR} = 1$ reflects the expected SNR. In other words, the expected SNR of a power-law GWB signal is encoded in its height (properly measured) above the reference PLI sensitivity curve for $\textrm{SNR} = 1$. 


\begin{figure}
\begin{center}
\includegraphics[width=0.67\textwidth]{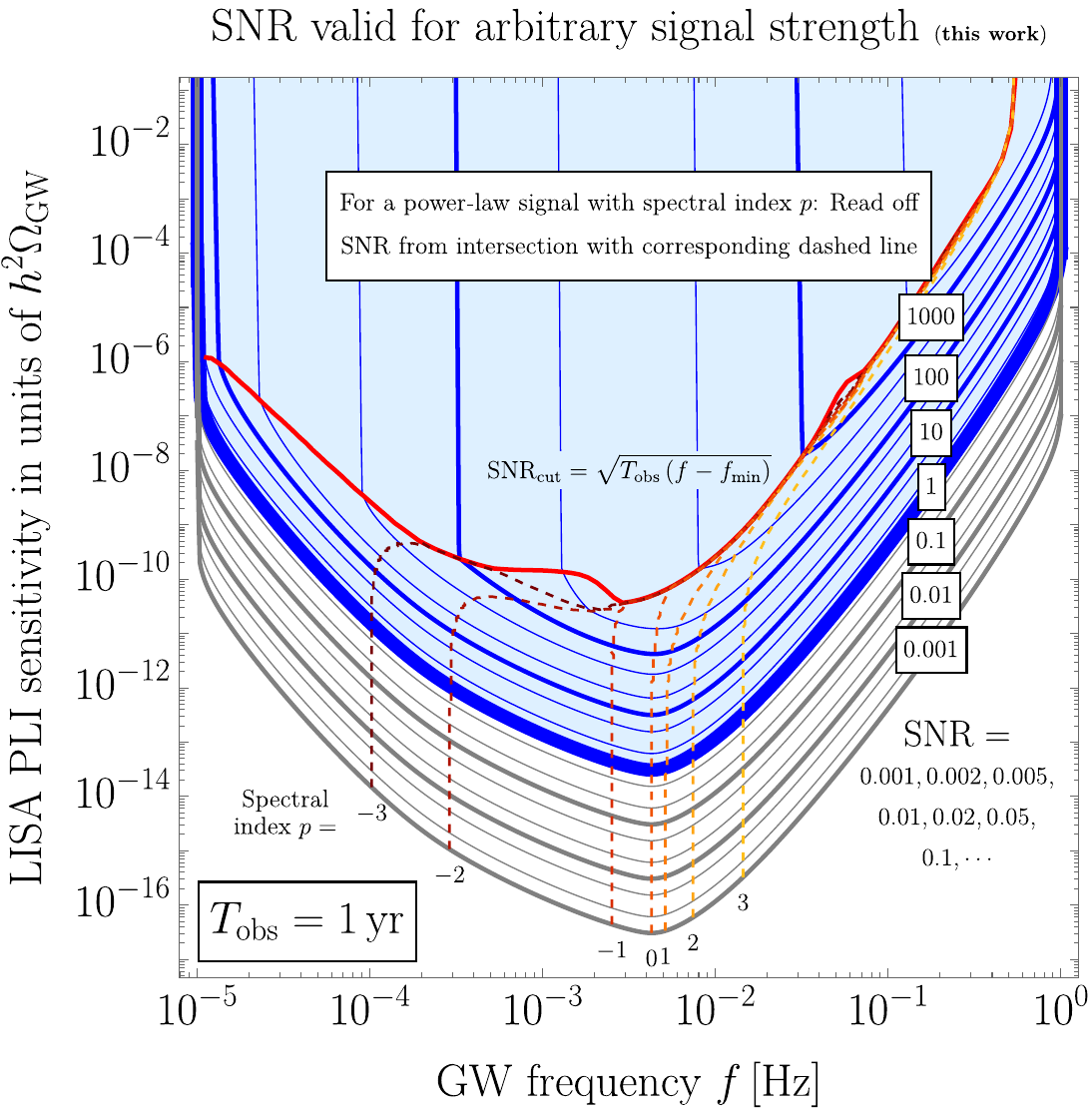}
\caption{PLI sensitivity curves derived from our new, fully general SNR expression (see Eq.~\eqref{eq:OPLIS}).}
\label{fig:PLISnew}
\end{center}
\end{figure}


After this short review of the standard construction of PLI sensitivity curves based on the expected SNR in the weak-signal regime, let us now discuss how the master formula in Eq.~\eqref{eq:SNRnew} can be used to construct PLI sensitivity curves that are valid for GWB signals of arbitrary strength. The philosophy behind this new, more general, construction will be exactly the same as before. Without loss of generality, we are able to write
\begin{equation}
\label{eq:OPLIS}
h^2\Omega_{\rm PLIS}(f;\textrm{SNR}) = \max_p \left[A_p(\textrm{SNR})\left(\frac{f}{f_{\rm ref}}\right)^p\right] \,.
\end{equation}
The important difference to the previous case, though, now lies in the computation of the amplitudes $A_p$, which requires us to solve a more complicated integral equation,
\begin{equation}
\label{eq:Ap}
\textrm{SNR} = \left[T_{\rm obs} \int_{f_{\rm min}}^{f_{\rm max}}{\rm d}f \:\left(\frac{A_p(\textrm{SNR})\,(f/f_{\rm ref})^p}{h^2\Omega_{\rm noise}(f)+A_p(\textrm{SNR})\,(f/f_{\rm ref})^p}\right)^2\right]^{1/2} \,.
\end{equation}
Obviously, solving this equation for $A_p$ in the limit of a weak signal results again in $A_p^{\rm WSR}$. In full generality, however, we need to solve Eq.~\eqref{eq:Ap} for $A_p$ numerically. The amplitudes thus obtained can then be used in Eq.~\eqref{eq:OPLIS} to revise the PLI sensitivity plot in Fig.~\ref{fig:PLISold}. The outcome of this exercise is shown in Fig.~\ref{fig:PLISnew}, which is a central result of our analysis.

The plot in Fig.~\ref{fig:PLISnew} differs from the one in Fig.~\ref{fig:PLISold} is several notable regards: 
(i) The PLI sensitivity curves are only parallel to each other in the limit of a weak signal. As we increase the signal strength, they first become distorted and then suddenly change their behavior completely. The information in the family of PLI sensitivity curves is therefore no longer redundant. Instead, each PLI sensitivity curve for a given SNR level carries relevant information. To emphasize this point, we will extend our terminology and refer from now on to the family of PLI sensitivity curves in Fig.~\ref{fig:PLISnew} as the \textit{LISA SNR contours}. 

(ii) As in Fig.~\ref{fig:PLISold}, we show again colored dashed lines in Fig.~\ref{fig:PLISnew} that allow one to read off the expected SNR for power-law GWB signals with spectral index $p=-3,-2,\cdots,2,3$. These curves, however, no longer extend only into the vertical direction. For GWB amplitudes around $h^2\Omega_{\rm signal} \sim 10^{-(11\cdots 9)}$, the dashed curves bend around, before they begin to follow the trend of a common attractor solution for $h^2\Omega_{\rm signal} \gtrsim 10^{-9}$. This behavior clearly reflects the transition from the weak-signal to the intermediate-signal regime and, subsequently, the situation deep inside the intermediate-signal regime. 

(iii) The sudden change in the behavior of the SNR contours at large GWB amplitudes and the attractor solution for the colored dashed lines in the intermediate-signal regime have a common origin: SNR contours in Fig.~\ref{fig:PLISnew} belonging to large SNR levels do not extend all the way from $f_{\rm min}$ to $f_{\rm max}$ but are restricted to a smaller interval $[f_{\rm cut},f_{\rm max}]$, where $f_{\rm cut}$ denotes the position of a low-frequency cutoff. To understand the origin of this cutoff, consider a power-law GWB signal with an asymptotically infinitely negative spectral index, 
\begin{equation}
\label{eq:Ospecial}
h^2 \Omega_{\rm signal}(f) = \lim_{p\rightarrow -\infty} A\left(\frac{f}{f_0}\right)^p \,,
\end{equation}
where the reference frequency, $f_{\rm ref} = f_0$, can now be freely chosen.
For this special choice of signal, the integrand in the SNR integral in Eq.~\eqref{eq:SNRnew} can be written as
\begin{equation}
\label{eq:thetaf0f}
\left(\frac{h^2\Omega_{\rm signal}(f)}{h^2\Omega_{\rm noise}(f)+h^2\Omega_{\rm signal}(f)}\right)^2 \quad\overset{p\rightarrow-\infty}{=}\quad \Theta\left(f_0-f\right) \,,
\end{equation}
which results again in a simple expression for the expected SNR (see also Eq.~\eqref{eq:SNRISR2}), 
\begin{equation}
\label{eq:SNRcut0}
\textrm{SNR} \quad\overset{p\rightarrow-\infty}{=}\quad \textrm{SNR}_{\rm cut}(f_0) = \sqrt{T_{\rm obs}(f_0-f_{\rm min})} \,. 
\end{equation}
By construction of our PLI sensitivity plot, this means that the special signal in Eq.~\eqref{eq:Ospecial} must be tangent to the SNR contour for SNR level $\textrm{SNR}_{\rm cut}$. But the special GWB signal in Eq.~\eqref{eq:Ospecial} will simply appear as a straight vertical line at $f=f_0$ in our plot. We thus conclude that the SNR contour for SNR level $\textrm{SNR}_{\rm cut}$ must diverge, $h^2\Omega_{\rm PLIS} \rightarrow \infty$, at
\begin{equation}
\label{eq:fcut0}
f_0 = f_{\rm min} + \frac{\textrm{SNR}_{\rm cut}^2}{T_{\rm obs}} \,.
\end{equation}
Finally, recall that $f_0$ can be arbitrarily chosen. The relations in Eqs.~\eqref{eq:SNRcut0} and \eqref{eq:fcut0} are thus fully general; we can either pick a frequency and compute $\textrm{SNR}_{\rm cut}$, or vice versa, 
\begin{equation}
\textrm{SNR}_{\rm cut}(f) = \sqrt{T_{\rm obs}(f-f_{\rm min})} \,, \qquad f_{\rm cut}(\textrm{SNR}) = f_{\rm min} + \frac{\textrm{SNR}^2}{T_{\rm obs}} \,.
\end{equation}

In Fig.~\ref{fig:PLISnew}, we show the SNR cutoff $\textrm{SNR}_{\rm cut}(f)$ as a function of $f$ as a solid red line. The interpretation of this SNR cutoff is straightforward: GWB signals that (i) exceed the noise by a large amount up to some frequency $f$ but that are (ii) strongly suppressed for all larger frequencies, can at most result in an SNR of $\textrm{SNR}_{\rm cut}(f)$. SNR contours for larger SNR levels therefore simply do not exist at frequencies below $f$. Similarly, one can relate $\textrm{SNR}_{\rm cut}(f)$ to $\textrm{SNR}_{\rm max}$ in Eq.~\eqref{eq:SNRSSR}. While $\textrm{SNR}_{\rm max}$ assumes a large, dominant GWB signal across the whole LISA frequency band $[f_{\rm min},f_{\rm max}]$, the SNR cutoff $\textrm{SNR}_{\rm cut}(f)$ does so only for the interval $[f_{\rm min},f]$. Correspondingly, we have $\textrm{SNR}_{\rm max} = \textrm{SNR}_{\rm cut}(f_{\rm max})$. 


\subsection{Cutoff contour}
\label{subsec:cutoff}


Let us now discuss the SNR cutoff in Fig.~\ref{fig:PLISnew} in even more detail. In the previous section, we have seen that an SNR contour for a given SNR level only extends over the restricted interval $[f_{\rm cut},f_{\rm max}]$. For $\textrm{SNR} = 100$, e.g., we find $f_{\rm cut} \approx 10^4\,\textrm{yr}^{-1} \simeq 0.32\,\textrm{mHz}$. At the same time, thanks to our numerical analysis, we also know the amplitude (in units of $h^2\Omega_{\rm GW}$) of each SNR contour at the frequency where it needs to be cut off, $h^2\Omega_{\rm PLIS}(f_{\rm cut};\textrm{SNR})$. In fact, the red cutoff contour in Fig.~\ref{fig:PLISnew} may be written as a function of $f$ as follows, 
\begin{equation}
\label{eq:Ocut}
h^2\Omega_{\rm cut}(f) = h^2\Omega_{\rm PLIS}(f;\textrm{SNR}_{\rm cut}(f)) = h^2\Omega_{\rm PLIS}(f;\sqrt{T_{\rm obs}(f-f_{\rm min})}) \,.
\end{equation}
From Fig.~\ref{fig:PLISnew}, it is apparent that $h^2\Omega_{\rm cut}$ looks strikingly similar to the LISA noise strain curve in Fig.~\ref{fig:noisePSD}; note, e.g., the bump in $h^2\Omega_{\rm cut}$ around $f \sim 10^{-3}\,\textrm{Hz}$, which looks similar to the bump in $h^2\Omega_{\rm noise}$ caused by galactic confusion noise. In this section, we shall now demonstrate that this similarity is not a coincidence, but that the $h^2\Omega_{\rm cut}$ and $h^2\Omega_{\rm noise}$ curves are indeed directly proportional to each other to good approximation. 

To begin with, let us pick again an arbitrary reference frequency $f_0$ deep inside the LISA frequency band, $f_{\rm min} \ll f_0 \ll f_{\rm max}$, and consider a power-law signal of the form
\begin{equation}
h^2\Omega_{\rm signal}(f) = A \left(\frac{f}{f_0}\right)^p \,.
\end{equation}
Then, for a steep negative spectral index, $p \ll -1$, the integrand of the SNR integral in Eq.~\eqref{eq:SNRnew} will be mostly sensitive to the behavior of LISA strain noise curve at frequencies close to $f_0$. For much smaller frequencies, $f \ll f_0$, the signal will strongly dominate over the noise, and the integrand will be close to $1$, while for larger frequencies, $f \gg f_0$, the signal will be suppressed in comparison to the noise, and the integrand will be close to $0$. We already encountered precisely this situation in Eq.~\eqref{eq:thetaf0f}, where we explicitly took the limit $p \rightarrow - \infty$. For now, however, let us keep the discussion more general and simply assume $p \ll -1$.  In this case, it will be a good approximation to describe the LISA strain noise curve simply in turns of a power law around the reference frequency $f_0$ as well,
\begin{equation}
h^2\Omega_{\rm noise}(f) \approx B \left(\frac{f}{f_0}\right)^q \,.
\end{equation}
To reiterate our point: this power law is supposed to capture the behavior of the noise curve at frequencies where $h^2\Omega_{\rm signal} \sim h^2\Omega_{\rm noise}$. At much lower or higher frequencies, the behavior of the noise may change; but this does not concern us, since, by construction, $h^2\Omega_{\rm signal} \gg h^2\Omega_{\rm noise}$ at lower frequencies and $h^2\Omega_{\rm signal} \ll h^2\Omega_{\rm noise}$ at higher frequencies, anyway, such that the actual shape of the noise curve does not matter there. So, from the perspective of the SNR, we may just assume the same power law for the noise throughout.

The power-law expressions for the signal and noise then allow us to write 
\begin{align}
\textrm{SNR} & \approx \left[T_{\rm obs} \int_{f_{\rm min}}^{f_{\rm max}}{\rm d}f \:\left(\frac{A\,(f/f_0)^p}{B\,(f/f_0)^q+A\,(f/f_0)^p}\right)^2\right]^{1/2} \\
& = \left[T_{\rm obs}f_0 \int_{x_{\rm min}}^{x_{\rm max}}{\rm d}x \:\left(\frac{x^p}{C\,x^q+x^p}\right)^2\right]^{1/2} \,,
\label{eq:xInt}
\end{align}
where, in the second step, we rescaled the frequency, $x = f/f_0$, and introduced $C = B/A$. The $x$ integral in Eq.~\eqref{eq:xInt} can be solved analytically in terms of a hypergeometric function,
\begin{equation}
\mathcal{I} = \int_{x_{\rm min}}^{x_{\rm max}}{\rm d}x \:\frac{1}{\left(1+C\,x^r\right)^2} = \left[{}_2 F_1\left(2,\frac{1}{r},1+\frac{1}{r},-C\,x^r\right)x\right]_{x_{\rm min}}^{x_{\rm max}} \,, \qquad r = q-p \,.
\end{equation}
For both $x_{\rm min} = f_{\rm min}/f_0 \ll 1$ and $x_{\rm max} = f_{\rm max}/f_0 \gg 1$, we may expand this hypergeometric function in the last, $x$-dependent argument. Keeping only the leading terms in these two expansions that are relevant for $p \ll -1$ (i.e., $r \gg 1$), we obtain
\begin{equation}
\mathcal{I} \approx C^{-1/r}\,\frac{\pi(r-1)}{r^2\sin(\pi/r)} - x_{\rm min} \,.
\end{equation}
Finally, in the last step, we expand this expression in the limit of a large spectral index $r$, 
\begin{equation}
\mathcal{I} \approx 1-x_{\rm min} - \frac{1+\ln C}{r}  \,,
\end{equation}
up to corrections of $\mathcal{O}(r^{-2})$. Putting everything together, we arrive at the rough estimate
\begin{equation}
\label{eq:SNRC}
\textrm{SNR} \approx \left[T_{\rm obs}\left(f_0-f_{\rm min}\right) - T_{\rm obs}f_0\,\frac{1+\ln C}{r} \right]^{1/2} \,,
\end{equation}


\begin{figure}
\begin{center}
\includegraphics[width=0.47\textwidth]{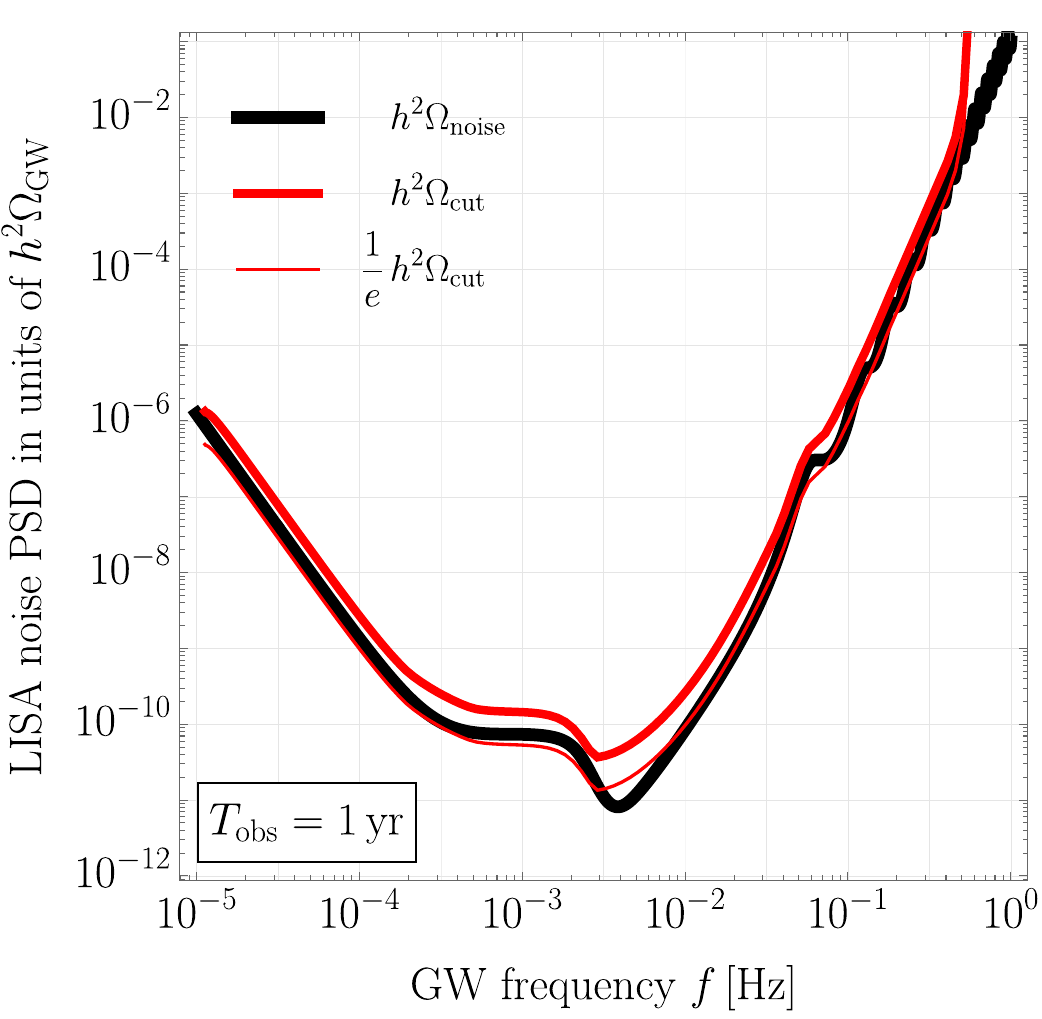}
\end{center}
\caption{Comparison between the LISA strain noise curve $h^2\Omega_{\rm noise}$ (see Fig.~\ref{fig:noisePSD}) and the cutoff contour $h^2\Omega_{\rm cut}$ in the LISA SNR contour plot (see Fig.~\ref{fig:PLISnew} and Eq.~\eqref{eq:Ocut}). Despite the many nontrivial steps in the construction of $h^2\Omega_{\rm cut}$, it turns out that $\frac{1}{e}\,h^2\Omega_{\rm cut} \sim h^2\Omega_{\rm noise}$ (see Sec.~\ref{subsec:cutoff}). }
\label{fig:Onoisecut}
\end{figure}


The estimate in Eq.~\eqref{eq:SNRC} allows us to draw some remarkable conclusions. First of all, we notice that the SNR cutoff $\textrm{SNR}_{\rm cut}$ is contained inside the expression in Eq.~\eqref{eq:SNRC}. Indeed, for $C = 1/e$, we recover exactly $\textrm{SNR}_{\rm cut}$, in accord with our expectation for power-law signals with a very large negative spectral index. At the same time, the choice $C=1/e$ renders the expression in Eq.~\eqref{eq:SNRC} independent of $r$, which can be interpreted as follows: consider a whole family of power-law GWB signals with large negative spectral indices $p$ whose amplitude at $f_0$ is larger than the amplitude of the noise curve at $f_0$ by a factor $A/B = 1/C = e$\,---\,equation~\eqref{eq:SNRC} then tells us that the signal curves belonging to this family all share approximately the same SNR, i.e., they are all approximately tangent to the same SNR contour. This situation describes exactly the behavior of the SNR contour for the respective SNR value around its low-frequency cutoff: first, the SNR contour bends quickly around (think of tangents with spectral indices $p=-10,-20,\cdots$), before it diverges to infinity, $h^2\Omega_{\rm PLIS}\rightarrow\infty$. In other words, we have now found that the cutoff of the SNR contour is located at $f=f_0$ and at an amplitude $h^2\Omega_{\rm cut} \sim e\,h^2\Omega_{\rm noise}$. But this statement must be true for any $f_0$ deep inside the LISA frequency band. We therefore conclude that the cutoff contour is roughly proportional to the LISA strain noise curve, $h^2\Omega_{\rm cut}(f) \sim e\,h^2\Omega_{\rm noise}(f)$. This conclusion is in excellent agreement with our numerical results for both curves, which are indeed approximately related by a factor of $e$ (see Fig.~\ref{fig:Onoisecut}). 


\subsection{Application: Gravitational waves from cosmic strings}


Having discussed the properties and new features of the LISA SNR contour plot in Fig.~\ref{fig:PLISnew}, let us finally apply the machinery developed in this paper to a concrete example. For this purpose, we shall pick a simple, one-parameter model that predicts a nonvanishing GWB spectrum across the entire LISA frequency band: GWs from a network of stable cosmic strings in the Nambu--Goto approximation. Specifically, we will work with the latest GWB templates for this type of strings, which also account for gravitational backreaction~\cite{Wachter:2024aos,Wachter:2024zly}. The only free parameter in this model is the dimensionless string tension $G\mu$, where $G$ is Newton's constant and $\mu$ the dimensionful string tension (i.e., energy per string length). 


\begin{figure}
\begin{center}
\includegraphics[width=0.47\textwidth]{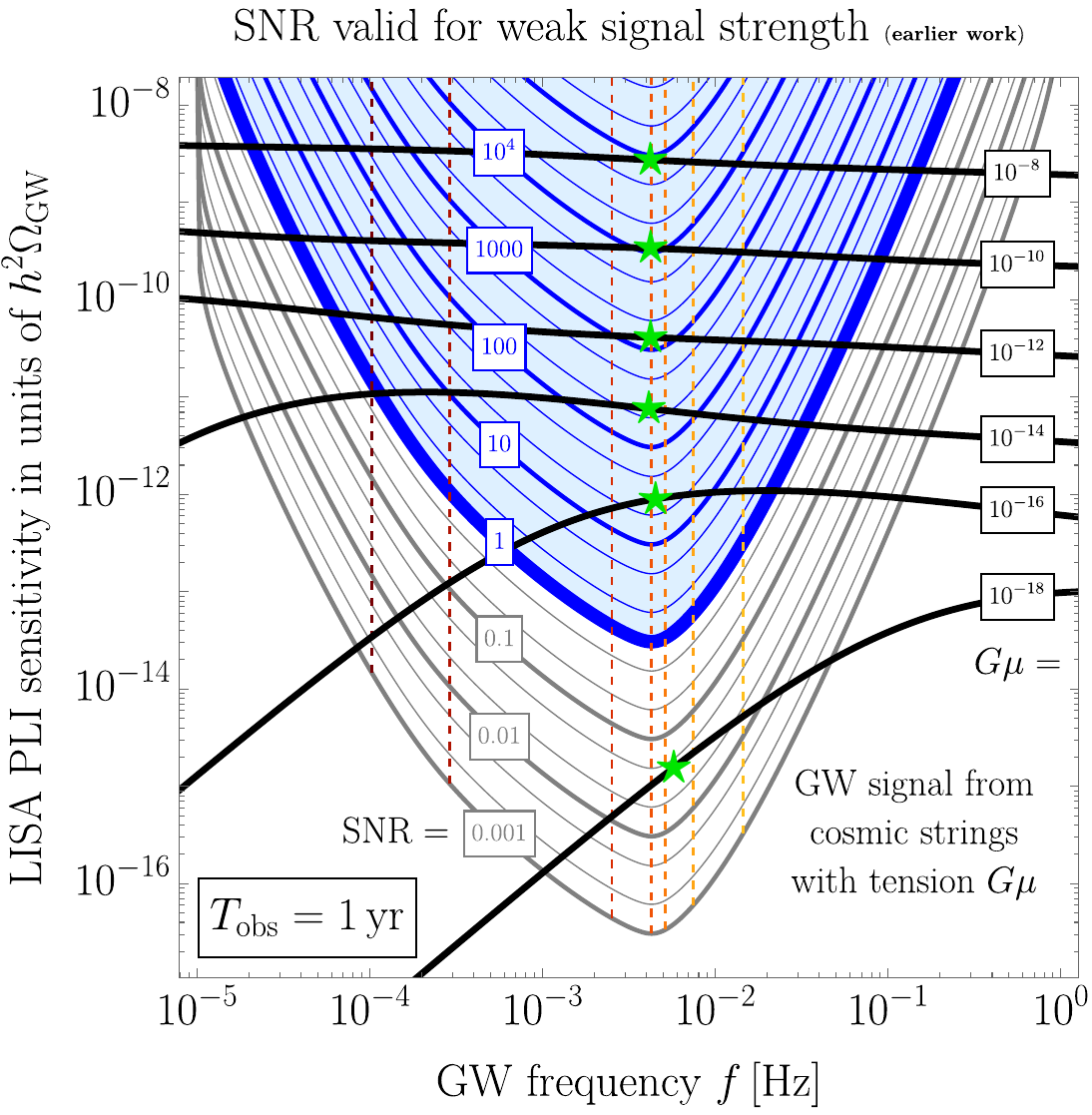}\qquad\includegraphics[width=0.47\textwidth]{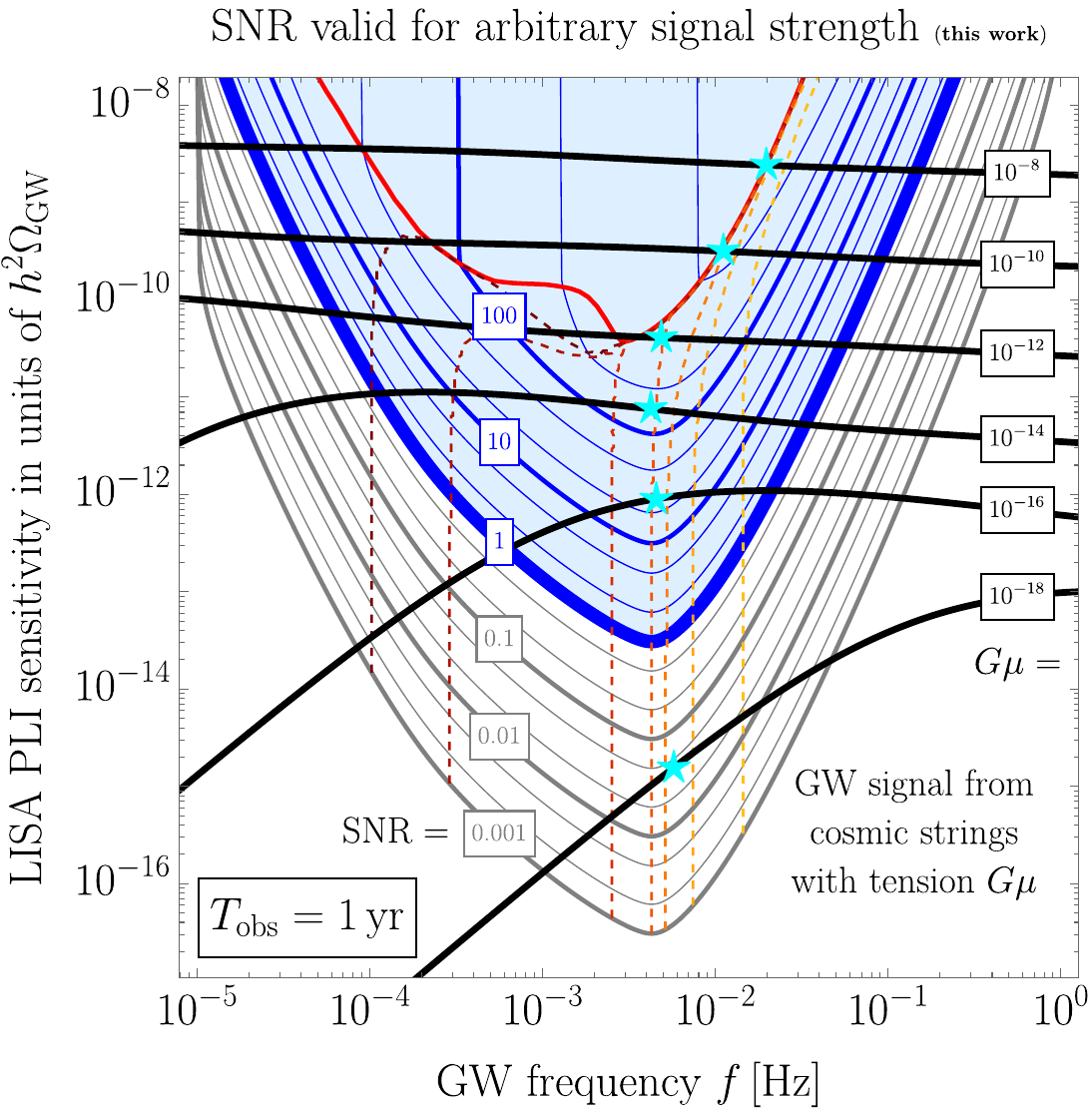}
\end{center}
\caption{GWB signals from cosmic strings for different values of the string tension on top of the old \textbf{(left panel)} and new \textbf{(right panel)} LISA SNR contour plots (see Figs.~\ref{fig:PLISold} and \ref{fig:PLISnew}). The green and cyan star-shaped markers indicate where to read off the SNR value for each GWB spectrum.}
\label{fig:string_spectra}
\end{figure}


In Fig.~\ref{fig:string_spectra}, we plot six representative GWB signals from cosmic strings on top of the old and new SNR contour plots that we already showed in Figs.~\ref{fig:PLISold} and \ref{fig:PLISnew}. In addition, we indicate with green and cyan star-shaped markers where to read off the SNR value for each GWB spectrum. In the case of the old SNR contour plot based on the SNR in the weak-signal regime (see Eq.~\eqref{eq:SNRold}), the graphical evaluation of the SNR is well known and straightforward: we simply need to find the SNR contour to which the signal curve is tangent\,---\,the SNR level of the SNR contour identified in this way then tells us the expected SNR for the GWB signal in question. In fact, the green markers indicate precisely the points along the signal curves where these curves are tangent to an SNR contour.

In the case of the SNR contour plot based on our new, general result for the SNR (see Eq.~\eqref{eq:SNRnew}), the graphical evaluation of the SNR is also straightforward, but involves one additional new rule:  we need to find the SNR contour to which the signal curve is tangent\,---\,if this is possible, nothing changes and the expected SNR can be read off as usual. But if it appears as if the signal curve is nowhere tangent to any SNR contour, we need to find the intersection between the signal curve and the red cutoff contour\,---\,the frequency where this happens, $f_0$, then determines the expected SNR, $\textrm{SNR} = \textrm{SNR}_{\rm cut}(f_0)$.

In fact, this new rule is not entirely necessary. At their low-frequency cutoff, the SNR contours simply bend around extremely fast, such that, after all, the new rule also just identifies the point on a signal curve where it is tangent to an SNR contour. For practical purposes, the red cutoff contour, however, provides useful guidance for the graphical evaluation of the SNR. The cyan markers in Fig.~\ref{fig:string_spectra} illustrate this new algorithm in a clear way: the first four makers (for $G\mu = 10^{-18},10^{-16},10^{-14},10^{-12}$) indicate where the respective signal curves are tangent to an SNR contour, while the last two markers (for $G\mu = 10^{-10}, 10^{-8}$) indicate where the respective signal curves intersect the red cutoff contour. 

\begin{figure}
\begin{center}
\includegraphics[width=0.47\textwidth]{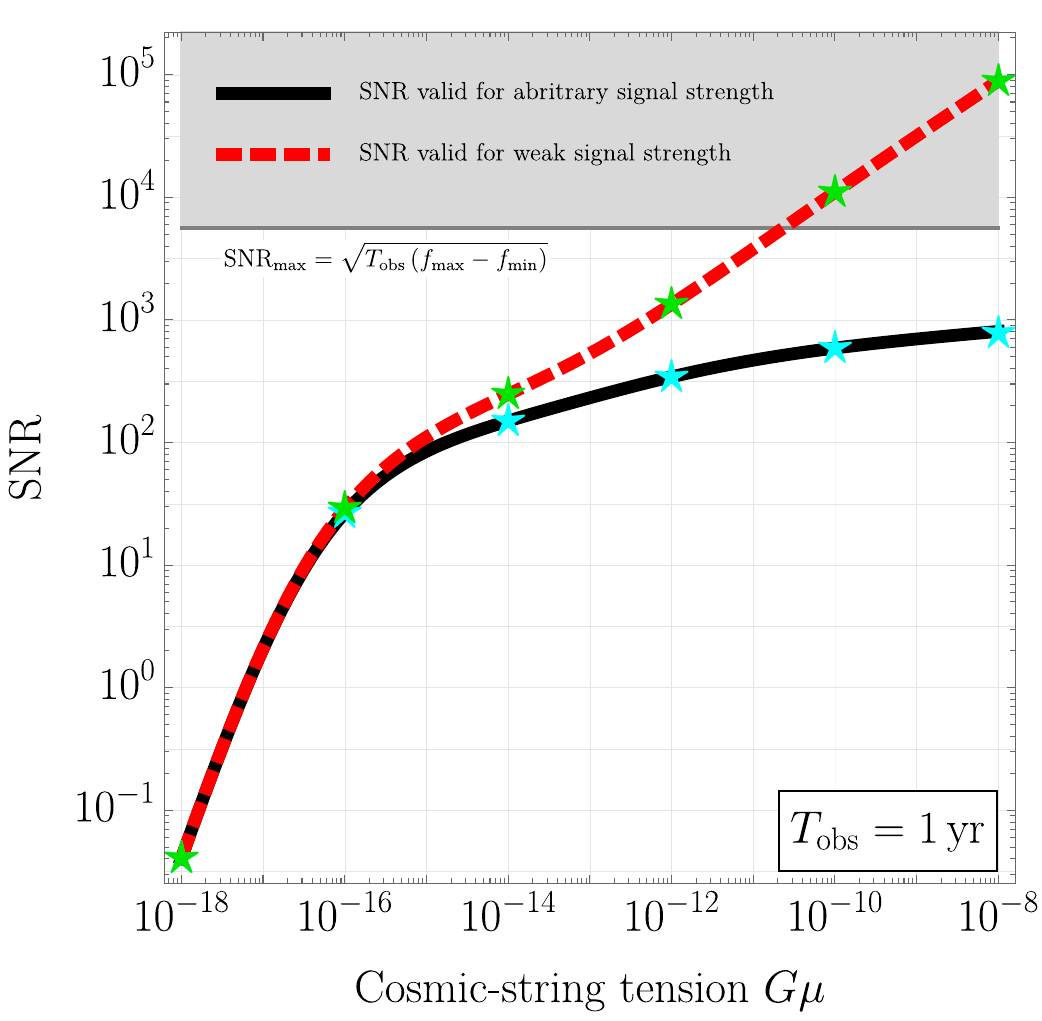}
\end{center}
\caption{Expected SNR for GWB signals from cosmic strings as a function of the tension $G\mu$. The dashed red line and the solid black line represent exact numerical results that follow directly from the expressions in Eqs.~\eqref{eq:SNRold} and \eqref{eq:SNRnew}, respectively, while the green and cyan star-shaped markers correspond to the graphical evaluation of the SNR based on the SNR contour plots in Fig.~\ref{fig:string_spectra}.}
\label{fig:string_snr}
\end{figure}

In Fig.~\ref{fig:string_snr}, we show the SNR values for the six GWB signals that can be read off from Fig.~\ref{fig:string_spectra}, again in terms of green and cyan star-shaped markers, and compare them to the exact numerical SNR values that follow directly from the expressions in Eqs.~\eqref{eq:SNRold} and \eqref{eq:SNRnew}. The agreement between both sets of SNR values is excellent\,---\,even though the GWB spectra from cosmic strings do not, in fact, correspond to constant power laws. This is a remarkable result, illustrating that the LISA SNR contour plot in Fig.~\ref{fig:PLISnew} can be used to graphically evaluate the expected SNR not only for pure power laws; it can also be used for GWB spectra that are approximated by a roughly constant power law whenever they are within the sensitivity reach of LISA (i.e., inside the blue-shaded region in Fig.~\ref{fig:PLISnew}). 

Another remarkable observation from Fig.~\ref{fig:string_snr} is that the standard expression for the SNR in Eq.~\eqref{eq:SNRold} only manages to accurately describe the expected SNR for string tensions up to $G\mu \sim 10^{-16}$. For larger tensions, one needs to work with our new expression in Eq.~\eqref{eq:SNRnew}. Doing so, one will then only find SNR values below $1000$ (considering string tensions of at most $G\mu = 10^{-8}$), which is in stark contrast to the inaccurate results derived from Eq.~\eqref{eq:SNRold}, which begin to exceed $\textrm{SNR}_{\rm max} \simeq 5600$ for tensions around $G\mu \sim 10^{-11}$. 


\section{Conclusions}
\label{sec:conclusions}


In this paper, we revisited the computation of the expected optimal SNR for a LISA auto-correlation measurement of a stochastic GWB signal. Earlier work had resulted in an expression for the SNR that is valid in the limit of a weak GWB signal (see  Eq.~(36) of Ref.~\cite{Thrane:2013oya}); here, we generalized this expression to arbitrary GWB signal strength. The important novelty of our new, general expression in Eq.~\eqref{eq:SNRnew} is that the total noise relevant for the computation of the SNR receives two contributions: the regular detector noise and a signal contribution that may be referred to as GWB self-noise. This GWB self-noise matters when one is interested in characterizing properties of the GWB signal (e.g., its SNR as a function of the signal amplitude) under the signal hypothesis. We specifically focused on the SNR for a LISA auto-correlation  measurement; the case of a cross-correlation measurement in a detector network had been discussed before in Ref.~\cite{Allen:1997ad}. 

Based on our new expression for the expected SNR, we were also able to generalize the construction of PLI sensitivity curves (see Figs.~\ref{fig:PLISold} and \ref{fig:PLISnew}), which led us to the notion of SNR contours. In the weak-signal limit, the PLI sensitivity curves are simply proportional to the expected SNR, $h^2\Omega_{\rm PLIS}(f) \propto \textrm{SNR}$, and thus do not contain any new information beyond the information that is encoded in the standard PLI sensitivity curve for $\textrm{SNR} = 1$. At larger signal strength, however, the situation changes, and each PLI sensitivity curve encodes unique information, which motivated us to extend our terminology and introduce the notion of SNR contours, i.e., PLI sensitivity curves for a fixed observing time but at different SNR levels (see also the title of the paper: \textit{SNR Contours for LISA}).

A remarkable feature of our SNR contour plot in Fig.~\ref{fig:PLISnew} consists in the presence of a cutoff contour marking the endpoints of individual SNR contours. We discussed in detail the origin of this cutoff, tracing it back to a frequency-dependent upper limit $\textrm{SNR}_{\rm cut}(f)$ for GWB signals that strongly dominate over the noise at all frequencies below $f$, but which are negligibly small for frequencies above $f$. For $f = f_{\rm max}$, we notably found the upper limit $\textrm{SNR} \leq \textrm{SNR}_{\rm max} \lesssim 10^4$ for typical LISA mission parameters. At the same time, we were able to understand the shape of the cutoff contour $h^2\Omega_{\rm cut}$; a rough analytical evaluation of the SNR in Eq.~\eqref{eq:SNRnew} allowed us to demonstrate that, approximately, $h^2\Omega_{\rm cut} \sim e\,h^2\Omega_{\rm noise}$. 

Finally, we applied the machinery developed in this paper to a concrete example: GWs from a network of stable cosmic strings in the Nambu--Goto approximation. This exercise led to several interesting insights. First of all, we were able to show that the SNR contour plot in Fig.~\ref{fig:PLISnew} can also be used for the graphical evaluation of the expected SNR if the GWB signal in question is not a perfect power law. This is an encouraging observation that significantly enlarges the range of applicability of the results obtained in the present paper. Second, we found that the standard SNR expression in the weak-signal regime only accurately describes the SNR for cosmic strings up to string tensions of around $G\mu \sim 10^{-16}$. For larger tensions, one necessarily needs to work with our new SNR expression. 

In summary, we conclude that the SNR expression in Eq.~\eqref{eq:SNRnew} and the SNR contour plot in Fig.~\ref{fig:PLISnew} represent powerful new tools that promise a wide range of applications in LISA sensitivity forecasts. It will be interesting to construct similar SNR contour plots for other detectors and detector networks in future work. 


\section*{Acknowledgments}

J.\,D.\,R.\ acknowledges support from (NSF) Grant No.\ PHY-2207270 and start-up funds at the University of Texas Rio Grande Valley.
K.\,S.\ is an affiliate member of the Kavli Institute for the Physics and Mathematics of the Universe (Kavli IPMU) at the University of Tokyo and supported by the World Premier International Research Center Initiative (WPI), MEXT, Japan (Kavli IPMU).
J.\,D.\,R.\ and K.\,S.\ are members of the NANOGrav Collaboration. NANOGrav is supported by NSF Physics Frontier Center award \#2020265.



\bibliographystyle{JHEP}
\bibliography{arxiv_1.bib}


\end{document}